# Synergistic Energy Absorption Mechanisms of Architected Liquid Crystal Elastomers

*Seung-Yeol Jeon[1,2,3, +], Beijun Shen[1,2, +], Nicholas A. Traugutt[4], Zeyu Zhu[1,2], Lichen Fang[1,2], Christopher M. Yakacki[4], Thao D. Nguyen[1,2,5], Sung Hoon Kang[1,2,5, *]*
(+: Equal contribution)

Dr. S.-Y. Jeon, B. Shen, Z. Zhu, Dr. L. Fang, Prof. T. D. Nguyen, Prof. S. H. Kang
Department of Mechanical Engineering
Johns Hopkins University
Baltimore, MD 21218, USA
E-mail: shkang@jhu.edu

Dr. S.-Y. Jeon, B. Shen, Z. Zhu, Dr. L. Fang, Prof. T. D. Nguyen, Prof. S. H. Kang
Hopkins Extreme Materials Institute
Johns Hopkins University
Baltimore, MD 21218, USA

Dr. S.-Y. Jeon
Institute of Advanced Composite Materials
Korea Institute of Science and Technology (KIST)
Wanju-gun, Jeonbuk, 55324, Republic of Korea.

Dr. N. A. Traugutt, Prof. C. M. Yakacki
Department of Mechanical Engineering
University of Colorado Denver
Denver, CO 80217, USA.

Prof. T. D. Nguyen, Prof. S. H. Kang
Center for Additive Manufacturing and Architected Materials
Johns Hopkins University
Baltimore, MD 21218, USA

**Abstract**: Here, we report the rate-dependent energy absorption behavior of a liquid crystal elastomer (LCE)-based architected material consisting of repeating unit cells of bistable tilted LCE beams sandwiched between stiff supports. Viscoelastic behaviors of the LCE material cause the energy absorption to increase with strain rate according to a power-law relationship, which can be modulated by changing the degree of mesogens alignment during synthesis. For a strain rate of 600 s$^{-1}$, the unit cell structure shows up to a 5 MJ/m$^3$ energy absorption density, which is two orders of magnitude higher than the same structure fabricated from Polydimethylsiloxane (PDMS), and is comparable to the dissipation from irreversible plastic deformation exhibited by denser metals. For a stacked structure of unit cells, viscoelasticity also produces nonuniform buckling of the LCE beams, causing the energy absorption density to increase with the stacking number *n* up to *n=3*. Varying the beam geometry further promotes the nonuniform buckling behavior allowing the energy absorption density to increase with stacking number without bounds. We envision that our study can lead to the development of lightweight extreme energy-absorbing materials.

Energy absorbing materials are used in many daily and advanced applications for comfort (e.g., in running shoes and vehicle engine mounts), performance (e.g., in sports equipment), and protection (e.g., in helmets and armor). These materials typically dissipate energy through inelastic mechanisms, such as plasticity and viscoelasticity, and through fracture and fragmentation. More recently, architected materials (or metamaterials) have been designed to trap energy through elastic buckling instabilities[1-5]. This structural energy trapping mechanism is scalable and reversible, which makes the architected material reusable[3]. Nevertheless, the elastic energy trapping mechanism is limited compared with other energy dissipation mechanism[6-8] because it has a fixed energy absorption capability regardless of strain rates[3]. It would be desirable to develop a reusable architected material that exhibits increased energy absorption with strain rate for enhanced impact protection performance.

To achieve the objective, we hypothesized that the capacity for energy absorption could be augmented further by combining the structural energy trapping mechanism with a nonlinear material dissipation mechanism[3,6-8]. While the concept of the architected material is based on the interplay between material and geometry, most of the studies were focused on nonlinear geometrical effects, and little has been studied about nonlinear material effects[5]. To test our hypothesis, we investigated exploiting the large viscoelastic dissipation of liquid crystal elastomers (LCEs) to develop reusable architected materials with tailorable rate-dependent enhancements to the energy absorption behavior. LCEs have a superior energy dissipation capability compared to amorphous elastomers. The material exhibits a high dissipation factor ($\tan\delta > 0.5$) over a broad range of frequencies and temperatures[9,10], and large hysteresis in the stress-strain response[11,13,23]. The hysteresis depends on the strain rate and the orientation of the director (i.e., the dominant orientation of the mesogens) relative to the loading axis and the order parameter (i.e., the degree of mesogen alignment).

We fabricated bistable architected structures consisting of repeating unit cells of symmetrically arranged tilted LCE beams sandwiched between two stiff horizontal supports. The energy absorption behaviors of stacked (multilayered) structures with different stacking numbers were measured by uniaxial compression tests under different strain rates. Over the range of effective strain rates from $6 \times 10^{-4} s^{-1}$ to $6 \times 10^{2} s^{-1}$, the energy absorption density of the LCE architected structure increased by more than two orders of magnitude with a power-law relationship. The rate of increase (i.e., the exponent of the power-law) can be modulated by changing the mesogen alignment, which is readily tunable by controlling the stretch applied during network formation. More remarkably, the energy absorption density increased with increasing stacking number, nearly doubling from a 1-layer to a 4-layer structure at the highest strain rate in the experiments. Finite element simulations of the stacked structures showed that viscoelastic behavior caused the layers to buckle sequentially, producing additional dissipation. However, for stacked structures with more than three layers of unit cells, the layers buckle more simultaneously, causing the energy absorption density to saturate at three layers. For higher stacking number structures, introducing variations in the LCE beam geometry induces sequential buckling of the layers and produces greater synergism in the viscous dissipation and structural energy trapping mechanisms.

To fabricate the bi-stable architecture, we first synthesized main-chain LCEs based on a two-stage thiol-acrylate reaction[12]. A diacrylate mesogen, RM257 (1,4-bis[4-(acryloyloxy)-propoxybenzoyloxy]-2-methylbenzene), was used to form the liquid crystal phase. To connect the mesogens, a di-functional monomer (C3, 1,3-propanedithiol) and a tetra-functional thiol monomer (PETMP, Pentaerythritol tetrakis (3-mercaptopropionate)) were selected as a flexible spacer and a crosslinker, respectively (**Figure 1a**, Supporting Information Section S1). During the first stage, a polydomain LCE was created by the thiol–acrylate Michael addition reaction. Next, the global arrangement of mesogens was further tailored by applying mechanical strain. After stretching a polydomain LCE, the photopolymerization reaction was used to establish crosslinks between the excess acrylate groups and fix the mesogen arrangement. Mechanical stretch causes the mesogens to rotate towards the loading direction and produce a soft stress response starting at 30% strain. At 90% strain, most of the mesogens have rotated towards the loading directions to complete the transformation of the polydomain to a monodomain[13]. To investigate the effect of mesogen alignment on energy absorption, we prepared LCE sheets with various mesogen alignments by applying different mechanical strains - 0% (random alignment) and 90% (full alignment) before photo-crosslinking. We also investigated the effect of the alignment direction by cutting beams from the sheets parallel or perpendicular to the director. The beams were tilted and inserted into the rigid body to create a bistable unit cell structure (**Figure 1b**, Supporting Information Section S2).

The unit cell structures with different mesogen alignment with respect to the beam direction were subjected to uniaxial compression at different effective strain rates, defined as displacement rate divided by the applied displacement, from $6\times10^{-4}$ s$^{-1}$ to $6\times10^{-1}$ s$^{-1}$ using an MTS Insight 5, from $8\times10^{-2}$ s$^{-1}$ to 8 s$^{-1}$ using TA Instruments ElectroForce 3200, and from $6\times10^{1}$ s$^{-1}$ to $6\times10^{2}$ s$^{-1}$ using Instron CEAST 9350 (**Figure 1c**, Movie S1). The energy absorption density was calculated as the area under the force-displacement curve normalized by the beam volume (**Figures 2a, 2b**, Supporting Information Sections S3, S4). The energy absorption density increased 10 times when the displacement rate increased from $10^{-2}$ mm/s to $10^{1}$ mm/s (effective strain rate: $6\times10^{-4}$ s$^{-1}$ - $6\times10^{-1}$ s$^{-1}$) (**Figure 2c**). Note that the LCE structure exhibits pseudo-bistability due to the viscoelastic behavior of the material[15,16]. This means that the structure maintains the snap-buckled configuration for a period of time then recovers the initial shape. The energy absorption density increased by two orders of magnitude when the effective strain rate increased from $6\times10^{-4}$ s$^{-1}$ to $6\times10^{2}$ s$^{-1}$ (**Figures 2c-2e**). For the range of effective strain rates from $6\times10^{-4}$ s$^{-1}$ to $6\times10^{-1}$ s$^{-1}$, the energy absorption density of the LCE structure increased with strain rate according to the power-law relationship, $E \propto \dot{\varepsilon}^{\alpha}$, where $\alpha = 0.323$. In comparison, the energy absorption density of the same unit cell structure fabricated using polydimethylsiloxane (PDMS) beams did not vary with strain rate (**Figure 2b**). For the highest strain rate in the MTS experiment, the energy absorption density of the polydomain unit cell was nearly an order of magnitude higher than that of the PDMS unit cell.

Next, we focused on the effect of mesogen alignment on tenergy absorption by using LCE samples with three different alignments with respect to the beam (longitudinal) direction. The energy absorption density of LCE structures for all cases followed a power-law relationship with the strain rate (**Figures 2c-e**, and **Table 1**). Beams fabricated from monodomains generally absorbed more energy than polydomain in relatively low strain rates. For the low strain rates of the MTS tests (**Figure 2c**), the polydomain and 90%∥ beams had similar power-law exponents, while the 90%⊥ beam had a substantially lower exponent. At the moderate strain rates of the ElectroForce tests (**Figure 2d**), the power-law exponent increased from polydomain through 90%⊥ to 90%∥. In contrast, during the high range of strain rates by the Instron Ceast experiments (**Figure 2e**), the energy absorption density was less sensitive to the strain rate and mesogen alignment.

To understand the rate-dependence of the energy absorption density of the LCE bistable beam structures, we characterized the master curve of the storage and loss moduli of the LCE materials with different mesogen alignments by the time-temperature superposition principle (**Figure 3a**, Figure S9)[17]. The variation in the power-law exponents for the rate-dependence of the energy absorption density of the unit cells with mesogen alignment can also be explained by the variation in the frequency dependence of the storage modulus. The strain rate in uniaxial compression tests is proportional to the frequency in dynamic mechanical analysis (DMA)[21]. At a few frequencies lower than $10^{-2}$ Hz, the storage modulus of the 90%∥ was larger than for the 90%⊥, and the power-law exponent abruptly decreased to 0.32 for the polydomain and to 0.14 for 90%⊥, respectively. Likewise, at the lower strain rate range ($6\times10^{-3}$ s$^{-1}$ - $6\times10^{-1}$ s$^{-1}$), the energy absorption density of the 90%∥ beam was larger than that of the 90%⊥. The energy absorption density of the 90%⊥ beams also had the flattest rate dependence in the low strain rate regime (**Figure 2c**). The storage and loss moduli for all materials showed little variation for high frequencies >$10^5$ Hz, indicating glassy behavior. This may explain why the energy absorption density of the LCE bistable beams was less sensitive to strain rates in the high-rate experiments (**Figure 2e**). Below $10^1$ Hz (**Figure 3a**), the storage modulus was larger for the monodomain than polydomain beams. The larger stiffness translated into greater energy absorption density in these comparisons (**Figures 2c and 2d**). For frequencies greater than 1 Hz but below $10^3$ Hz, the storage moduli for the polydomain and 90%⊥ decreased with decreasing frequency with a power-law exponent of about 0.5 (**Figure 3a**), which is characteristic of the Rouse model. In contrast, the storage modulus for the 90%∥ had a constant power-law exponent of 0.32. The mesogen alignment along the loading direction may inhibit the chain dynamics leading to the constant frequency dependence. Since deformation parallel to the director does not induce mesogen rotation, we speculate that the change in the frequency dependence of the modulus observed in the polydomain and 90%⊥ but not in the 90%∥ was caused by a change in dissipation mechanism from viscous chain dynamics to viscous mesogen rotation. Previous works showed that the nematic director relaxation time is longer than the characteristic Rouse time[20]. Therefore, viscous mesogen rotation is likely responsible for the smaller power-law exponent in the frequency dependence of the storage modulus at a lower frequency (**Figure 3a**) and in the rate-dependence of the energy absorption density at lower rates for the 90%⊥ beam (**Figure 2c**). The effect of mesogen alignment on the molecular relaxation mechanism is summarized schematically in **Figure 3b**.

The relaxation spectrum was determined for the different materials from the master curve of the storage moduli (Figure S10) and applied to finite element simulations of the rate-dependent compression experiments (**Figure 3c**, Supporting Information Sections S6-S8). A finite deformation viscoelastic model[25] with a discrete relaxation spectrum was used to describe the behavior of the LCEs. The contributions to energy absorption from the stored energy and the viscous dissipation were computed for the viscoelastic model (Supporting Information Section S7). Both the stored and dissipated energy density increased with strain rate for all materials for the strain rates $8.3\times10^{-4}$ s$^{-1}$ – $8.3\times10^2$ s$^{-1}$ in experiments. Both the stored and dissipated energy density of the polydomain is smaller than those of 90%∥ and 90%⊥ over the entire range of strain rates, which is consistent with the frequency dependence of the master curve of the storage modulus of the 3 LCEs (**Figure 3a** and Figure S9). During the low-rate regime corresponding to MTS test in **Figure 2c**, the stored energy density for all three materials increased less than one order of magnitude, whereas the viscous dissipation density increased more than an order of magnitude. Compared to stored energy, viscous dissipation is more sensitive to strain rate during the low strain rate range, regardless of mesogen alignment (**Figure 3c**). In contrast, as strain rate increased, the rate-dependence of viscous dissipation for 90%⊥ became flatter, but that of 90%∥ did not change. It is likely because in the 90%⊥ case, there is also a mesogen rotation dissipation mechanism, whereas in 90%∥, there is only viscous chain dynamics. Due to the increasing contribution of stored and dissipated energy (**Figure 3c**), the total energy absorption increased with strain rate (**Figures 2c-e**).

We next investigated the energy absorption of multi-layered bistable structures by arranging the unit cells with polydomain beams in $2\times n$ arrays, where $n$ is the stacking number (Supporting Information Section S2). **Figure 4a** shows a series of snapshots of a $2\times2$ structure under uniaxial compression at an effective strain rate of $2.38\times10^{-2}$ s$^{-1}$ (Movie S2). Both layers are initially compressed together. However, instead of both layers buckling simultaneously, the top layer began to buckle, and the second layer recovered (i.e., straightened). As the top layer collapsed, the bottom layer again compressed and buckled. The sequential buckling of the layers is also evident in the plot of the force-displacement curve (**Figure 4b)** that shows two distinct peaks. The larger first peak corresponds to the initial compression of the two layers and buckling of the top layer, while the second smaller peak marks the buckling of the bottom layer.

We observed the same sequential buckling behavior at the same strain rate for taller 2×3 and 2×4. The force-displacement curve had 3 and 4 peaks for the 2×3 and 2×4 structures (Figures S7 and S8). The sequential buckling of the layers caused the energy absorption density to increase with the stacking number *n* (**Figure 4c**). This effect was not observed for the multilayered structure with PDMS beams. The energy absorption density of stacked LCE structures also followed a power-law relationship with the strain rate (**Table 2**).

To better understand the mechanisms underlying the increase in the energy absorption density with the stacking number, we simulated the compression tests and calculated the rate-dependence of the stored energy density and viscous dissipation density for different stacking numbers (Supporting Information Section S7 and S8). The stored energy density and dissipation energy density are plotted as a function of the stacking number for the low strain rate of $2.27 \times 10^{-2}$ s$^{-1}$ in **Figure 4d**. As the stacking number increased, the dissipated energy density increased, while the stored energy density remained constant (as seen from **Figure 4d** and Figure S19). The increased dissipation density occurred because each unit cell experienced the process of compression, recovery, re-compression before buckling. The viscoelastic behavior introduced sufficient non-uniformity in the stress-state of the beams to produce non-uniform buckling, which then produced additional dissipation and increased energy absorption density. For taller stacked structures with stacking numbers more than 3, the unit cells in adjacent layers again buckled together, resulting in a smoother force-displacement curve, with fewer distinct peaks than the stacking number and the energy absorption density to plateau (Figures S16, S24).

To promote this dissipation mechanism for higher stacking number structures, we varied the thickness of the LCE beams in each layer to ensure sequential buckling of the different layers (Movie S3). The beam thicknesses were gradually increased from the top to bottom layers while maintaining the total volume of the beams the same as the uniform thickness cases. Specifically, for a 2×8 structure going from top to bottom layers, the beam thickness of the adjacent layer was 0.05 mm smaller than the preceding layer. The graded structure buckled sequentially from the top to the bottom, producing eight distinct peaks in the force-displacement curve (**Figure 4e**).

To quantitatively compare graded and uniform structures, we calculated the difference in the energy absorption density between graded and uniform cases at different strain rates (**Figure 4f**). The difference in the energy absorption density increased exponentially with strain rates. At the highest strain rate in **Figure 4f**, the energy absorption density of the 2×8 graded structure was 45% higher than the 2×8 uniform structure. Furthermore, the energy absorption density of graded stacked structures does not plateau at higher stacking numbers (Figure S26). The synergistic energy absorption mechanisms provide new opportunities for the design of the material architecture to optimize energy absorption.

In summary, we found that the energy absorption density of LCE-based metamaterials is significantly enhanced following a power-law relationship with strain rate. The power-law exponent is tunable by altering the degree of mesogen alignment. More mesogen alignment inhibits greater chain mobility, resulting in a flatter rate-dependence. Based on this newly acquired knowledge, we achieved over two orders of magnitude increase in the energy-absorbing capability of our LCE-based unit cell as the strain rate increased from $6 \times 10^{-4}$ s$^{-1}$ to $6 \times 10^{2}$ s$^{-1}$.

Additionally, we found that vertically stacking LCE units with uniform beam thickness caused nonuniform buckling, which increased the viscous dissipation density, which further increased the energy absorption density up to a factor of 1.6 for a stacking number of 4, compared to unit cells. This was not observed in the same structures made from PDMS. Further increases in the energy absorption density were realized by varying the thicknesses of the LCE beams from top to bottom to ensure sequential buckling. The energy absorption density of the 2×8 graded structure was 45% higher than that of the 2×8 uniform structure.

In this work, we used LCE as a beam material for a simple one-dimensional meta-structure, but further enhancement of energy absorption is expected through systematic structural design in the future. We envision our findings contribute to not only the fundamental understanding of the nonlinear energy absorption mechanisms of architected LCE structures but also applications where lightweight and extreme energy absorption are desirable, such as aerospace and automotive vehicles and personal protection.

**Supporting Information**
Supporting Information is available from the Wiley Online Library or the author.

**Acknowledgments**
S. J. and B. S. contributed equally to this work. This research was supported by the Army Research Office (Grant Number W911NF-17-1-0165) and the Johns Hopkins University Whiting School of Engineering start-up fund. The authors would like to thank Mr. Zheliang Wang for help with simulation using Tahoe and thank Dr. Brandon Zimmerman and Mr. Bibekananda Datta for helpful suggestions. The views and conclusions contained in this document are those of the authors and should not be interpreted as representing the official policies, either expressed or implied, of the Army Research Office or the U.S. Government. The U.S. Government is authorized to reproduce and distribute reprints for Government purposes notwithstanding any copyright notation herein.

Received: ((will be filled in by the editorial staff))
Revised: ((will be filled in by the editorial staff))
Published online: ((will be filled in by the editorial staff))

**Table 1.** Power-law exponent values of LCE structures with different mesogen alignments.

|  |  | Polydomain | 90%⊥ | 90%∥ |
|---|---|---|---|---|
| MTS Insight 5 | n | 0.323 | 0.217 | 0.336 |
|  | $R^2$ | 0.983 | 0.992 | 0.900 |
| TA ElectroForce 3200 | n | 0.261 | 0.312 | 0.348 |
|  | $R^2$ | 0.992 | 0.996 | 0.982 |
| Instron CEAST 9350 | n | 0.116 | 0.357 | 0.117 |
|  | $R^2$ | 0.920 | 0.911 | 0.984 |

**Table 2.** Constants of the power-law fit equation, $E = a\dot{\varepsilon}^b$, for uniform stacked LCE structures (Figure 4c).

| Stack number | $a$ | $b$ | $R^2$ |
|---|---|---|---|
| 1 | 0.63 x 10$^5$ | 0.198 | 0.925 |
| 2 | 1.04 x 10$^5$ | 0.264 | 0.999 |
| 3 | 1.01 x 10$^5$ | 0.231 | 0.945 |
| 4 | 1.28 x 10$^5$ | 0.249 | 0.962 |

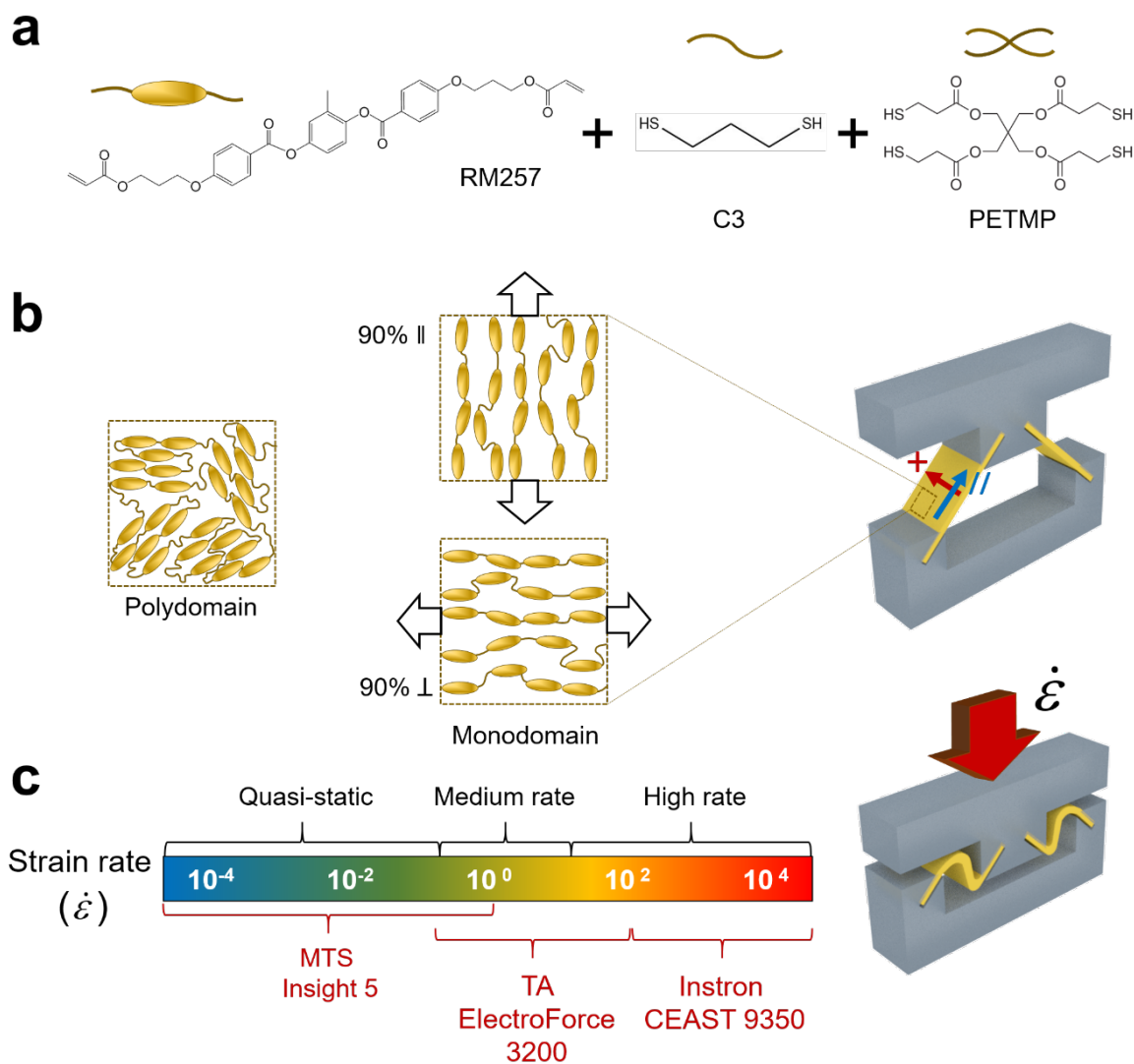

**Figure 1.** Mechanical energy absorption via bistable architected LCEs. a) LCEs are synthesized by a Michael addition reaction between diacrylate mesogen (RM257), dithiol spacer (C3), and tetrathiol crosslinker (PETMP). A non-equimolar excess of 2 mol% acrylate groups is used to fix mesogen arrangement by additional photo-crosslinking. b) Different mesogen arrangements in the bistable LCE beam are considered by applying different mechanical strains (0% and 90%). The effect of the direction of mesogen arrangement according to the mechanical strain is also investigated by inserting mesogen-arranged LCEs perpendicular (⊥) with or parallel (∥) to the beam direction. c) The energy-absorbing capability of LCE structures is characterized over a wide range of strain rates. Three different mechanical testing systems are introduced for an understanding of their strain-rate dependent energy absorption behaviors from quasi-static conditions ($10^{-4}$ s$^{-1}$~$10^{-1}$ s$^{-1}$) to dynamic conditions up to near $10^4$ s$^{-1}$.

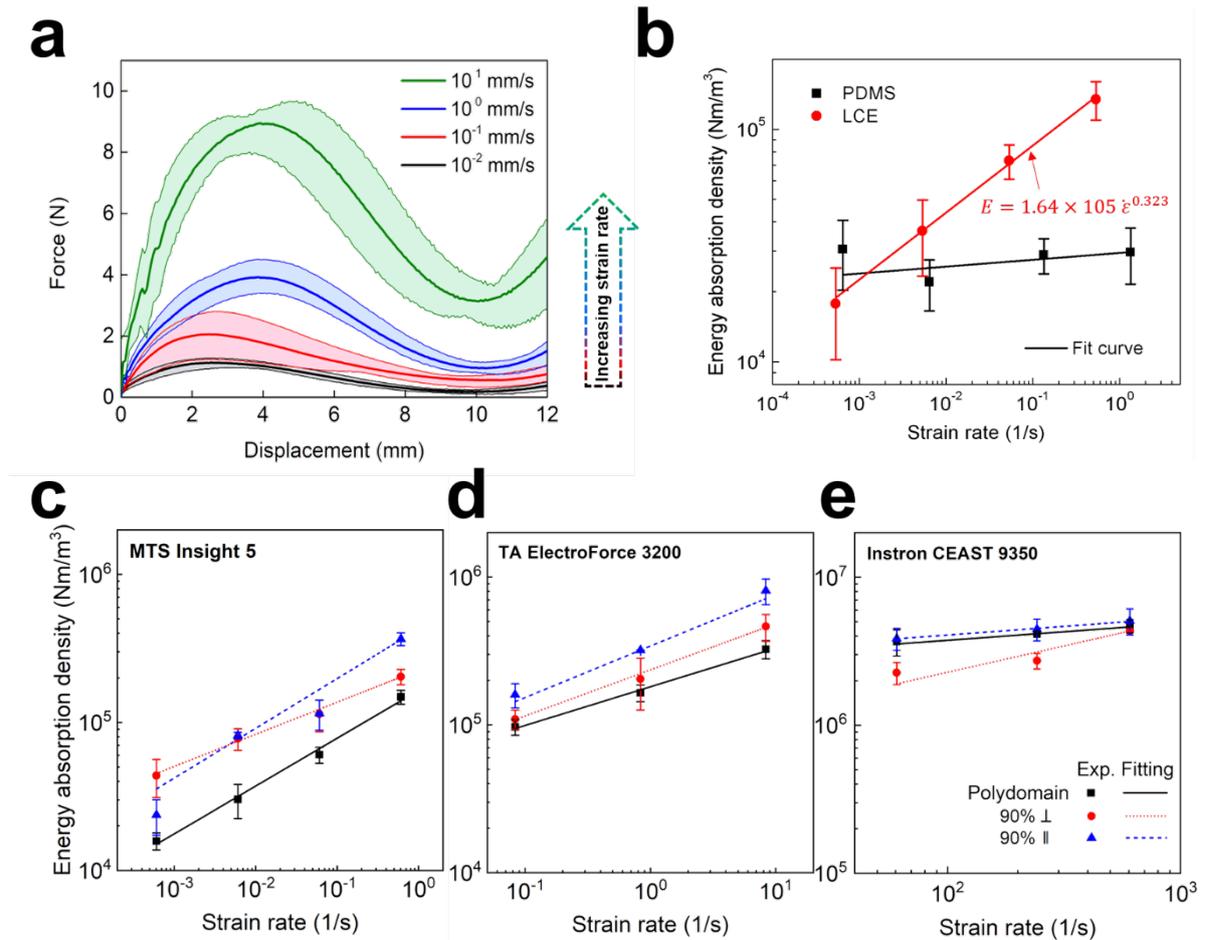

**Figure 2.** Mechanical energy absorption for a unit cell. a) Force–displacement curves for architected polydomain LCEs at multiple displacement rates. The area underneath a force-displacement curve indicates the energy absorption. The shaded regions indicate the maximum and minimum of 3 specimens. The solid line indicates the average value. b) Energy absorption density of architected polydomain LCE vs. PDMS. The energy absorption density follows the power-law relation (represented in red line) for strain rate ($\propto \dot{\varepsilon}^{0.323}$, $R^2$=0.983). The identical structure composed of polydimethylsiloxane (PDMS) is tested for comparison. c, d, e) Energy absorption densities of architected LCEs consisting of differently arranged liquid crystal molecules. The energy absorption is characterized by using MTS Insight 5 (c), TA ElectroForce 3200 (d), and Instron CEAST 9350 (e). The lines are the power-law fit between energy absorption and strain rate.

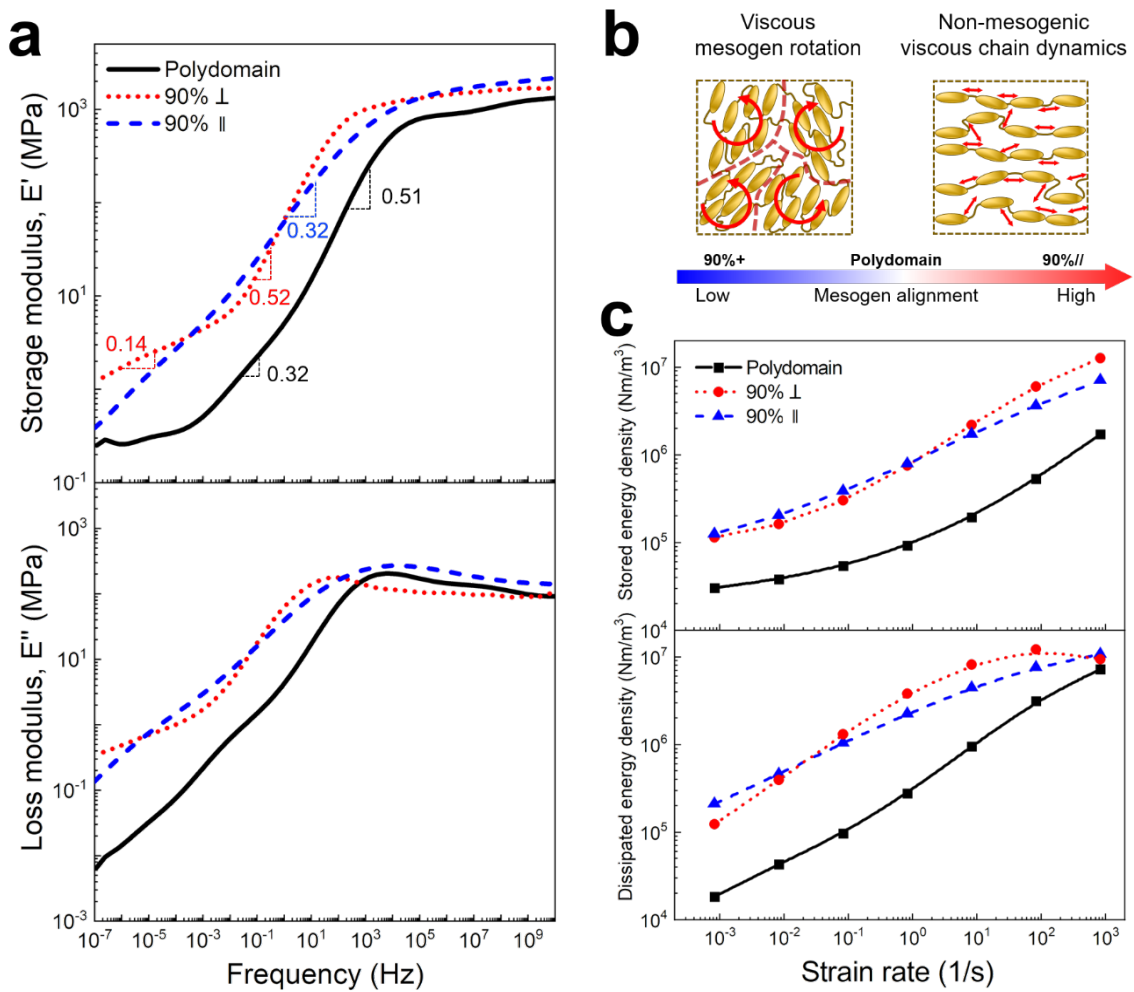

**Figure 3.** Mechanical energy absorption mechanisms according to the mesogen alignment (Polydomain, 90%⊥, 90%∥). a) The storage modulus (E') and loss modulus (E") as a function of frequency for LCEs with different mesogen alignments with respect to the beam direction. b) A schematic of relaxation mechanisms of LCEs depending on the mesogen alignment. c) The stored and dissipated energy contributions to total energy absorption calculated by finite element analysis using viscoelastic material properties determined from the master curve of the storage modulus of the 3 LCE materials.

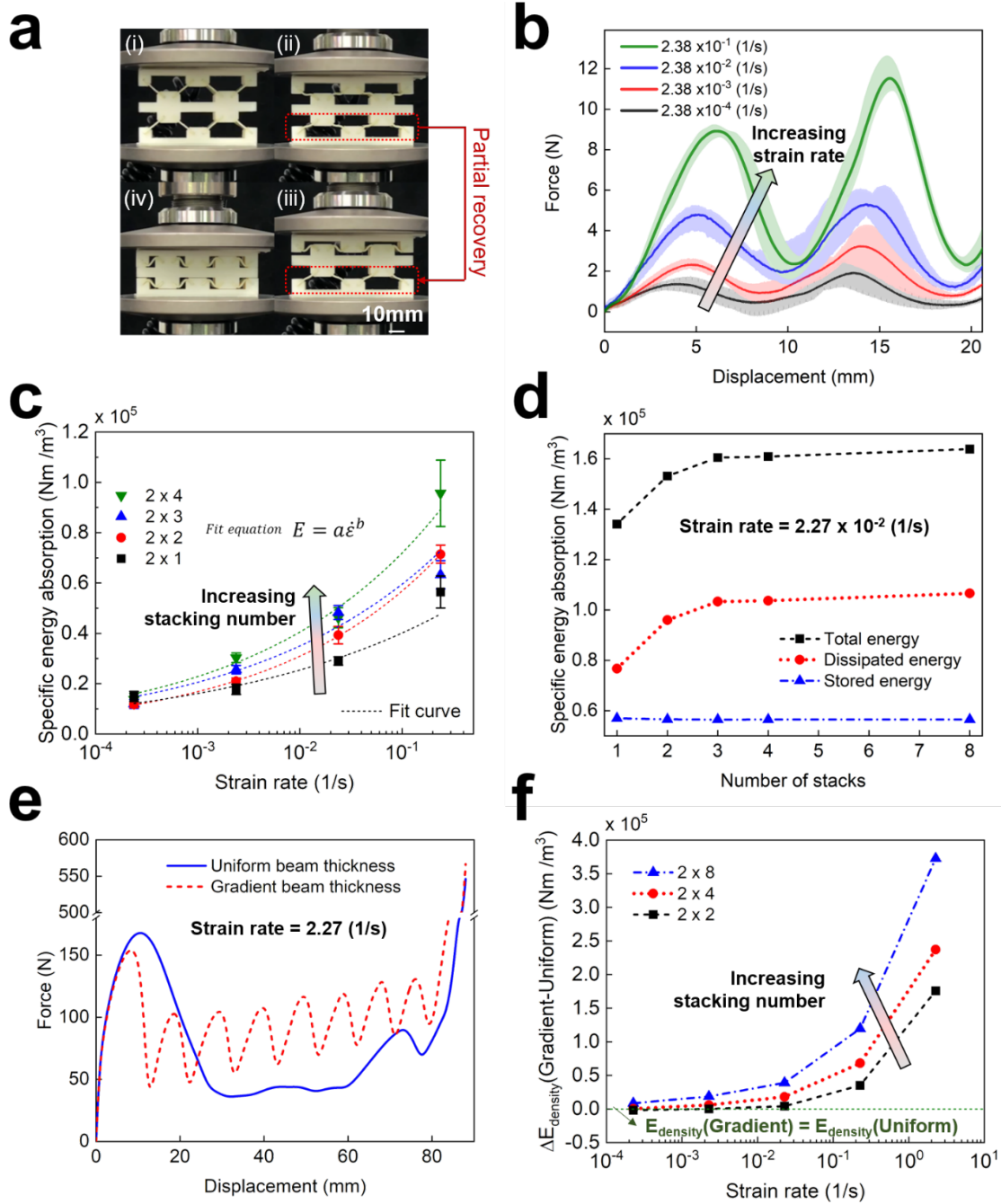

**Figure 4.** The mechanical energy absorption through stacked polydomain LCE structures. a) Experimental snapshots of a 2×2 stacked LCE structure under compressive loading at a strain rate of $2.38\times10^{-2}$ s$^{-1}$: (i)-(ii) initially both layers collapse simultaneously, (ii)-(iii) the bottom layer slightly recovers back (i.e., straightens) while the top layer keeps collapsing, (iii)-(iv) both layers collapse till fully deformed. b) The experimental force-displacement curves of 2×2 stacked LCE structures at different strain rates. The shaded region indicates the range of experimental measurement for at least three specimens. The solid lines indicate the average values. c) The energy absorption density of stacked LCE structures plotted as a function of strain rates for different stacking numbers from 2×1 to 2×4 (experiment). d) The total energy absorption density, dissipated energy density, and stored energy density of stacked LCE structure plotted as a function of stacking number at a strain rate of $2.27\times10^{-2}$ s$^{-1}$ (simulation). e) The force-displacement curves of 2×8 stacked LCE structures with uniform and graded beam thickness for a strain rate of 2.27 s$^{-1}$ (simulation). f) The difference in the energy absorption density between graded and uniform cases for 2×2, 2×4, and 2×8 stacked LCE structures plotted as a function of strain rates (simulation). The horizontal dash line passing through zero indicates the case where the energy absorption density of the structures with graded beam thickness is equal to those with uniform beam thickness.

# Supporting Information

**Synergistic Energy Absorption Mechanisms of Architected Liquid Crystal Elastomers**

*Seung-Yeol Jeon+, Beijun Shen+, Nicholas A. Traugutt, Zeyu Zhu, Lichen Fang, Christopher M. Yakacki, Thao D. Nguyen, Sung Hoon Kang\**

+: Equal contribution

**Table of Contents**
1. LCE synthesis
2. Sample fabrication
3. Force-displacement measurements
4. Force-displacement data for unit cell
5. Force-displacement data for stacked structures
6. Viscoelastic materials characterization
7. Viscoelastic model
8. Finite element models
9. Movies

**1. LCE synthesis**
*Materials*
    4-Bis-[4-(3-acryloyloxypropypropyloxy) benzoyloxy]-2-methylbenzene (RM257) was purchased from Wilshire Technologies, Inc. (Princeton, NJ, USA). Pentaerythritol tetra(3-mercaptopropionate) (PETMP), 1,3-Propanedithiol (C3), 2,6-di-tert-butyl-4-methylphenol (BHT), Phenylbis(2,4,6-trimethylbenzoyl) phosphine oxide (PPO), Triethylamine, and toluene were purchased from Sigma-Aldrich, Inc. (St. Louis, MO, USA). All materials were used as received without further purification.

*Fabrication*
    Main-chain liquid crystal elastomers were synthesized by a method adapted from the literature[1]. The stoichiometry of the monomer mixture was systematically calculated based on the thiol-acrylate click chemistry. The amount of crosslinker (PETMP) (13 mol.%) was calculated by the number of thiol functional groups contributed by crosslinker relative to the total number of thiol groups from crosslinker and spacer (C3).

    RM 257 and inhibitor BHT were added in a glass vial and mixed with a vortex mixer, followed by a dissolution in 30 wt% toluene. The solution was placed in an oven (Lindberg/Blue M™ Vacuum Oven, Thermo Fisher Scientific Inc, Waltham, MA, USA) at 90 °C for around 30 minutes. PETMP, C3 and PPO were added to the solution in sequence, and the mixture was put in the oven again at 100 °C for 3 minutes. Triethylamine was added to the mixture to initiate a thiol-acrylate Michael addition reaction of equal parts thiol and acrylate functional groups. The mixed solution was then placed in a desiccator to remove any air bubbles caused by the mixing process. Afterwards, the solution was cast between two glass slides with 1 mm polytetrafluoroethylene spacers and was allowed to cure for around 2 days under room temperature. During the time, the mixture polymerized, forming a clear elastomer sheet containing toluene. The toluene contained in the sheet was then completely extracted under 21 in-Hg vacuums at 90 °C for another 2 days.

For making polydomain LCEs, the cured LCE sheet was taken out from the oven and put into a UV chamber (CL-1000 Ultraviolet Crosslinker, Analytik Jena US LLC, Upland, CA, USA) and both sides were exposed to 365 nm wavelength ultraviolet light for up to 3 hours, respectively, during which excess acrylate groups reacted to form new bonds in the network. The resulting sheet was opaque white at room temperature, indicating a polydomain structure. To obtain specimens with aligned mesogens, the cured LCE sheet was stretched uniaxially to a strain of 90% before UV polymerization. The initially opaque sheet became transparent during stretching from the ordering of the mesogens. The stretched sheet as exposed to UV, at the same wavelength and intensity as was applied for the polydomain, on both sides to fix the alignment.

## 2. Sample fabrication

*Unit cell*

The unit cell samples were prepared with the LCE beams and rigid support structures by the interlocking assembly method. Synthesized LCE sheets were cut into a series of beams of length 14 mm, width 10 mm, and thickness 1.8 mm. The rigid support structures were fabricated using polylactic acid (PLA) (Polymaker, Shanghai, China, PolyLite PLA, 2.85 mm, Young's Modulus: 2636 ± 330 MPa) via a Fused Filament Fabrication (FFF) 3D printer (LulzBot TAZ 6, Fargo, ND, USA). CAD models of the support structures were created in SolidWorks then exported as an STL file. The SLT file was imported into Cura for slicing and to generate the G-CODE file for the 3D printer. To assemble the unit cell structure (Figure S1a), the LCE beams were glued to the support structures in secured grooves using a cyanoacrylate glue (Gorilla super glue, Cincinnati, OH, USA). For the LCE beams with pre-strained of 90%, depending on the pre-strain direction was either parallel or perpendicular to the beam direction, unit cells with 90%⊥ and 90%∥ LCE beams were made.

For the impact test, SAE 304 stainless steel was used to manufacture the bottom support (Figure S1b) to minimize deformation of the support frame. In addition, the design of the upper support was changed from a long bar to a short square bracket to reduce the mass. The top surface was covered with protective tape (3M 8641 Polyurethane Protective Tape).

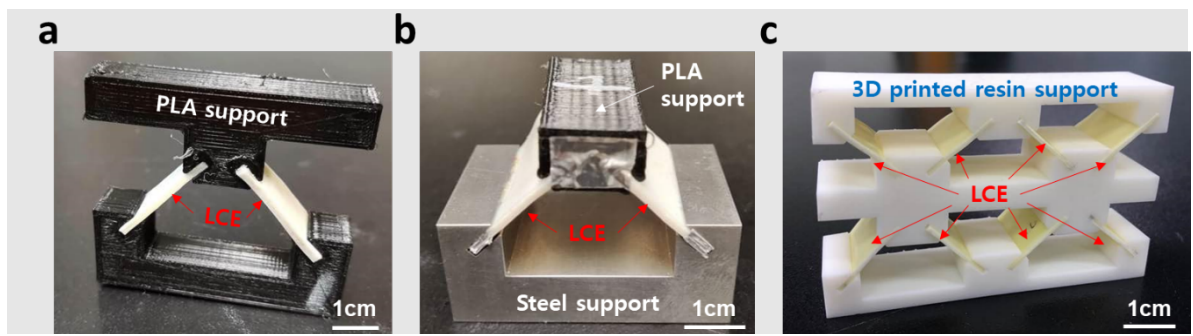

Figure S1. Unit cells prepared for compression tests at low and moderate strain rates (a) and for impact tests (b). (c) An example of a stacked structure (2×2 stacked structure).

*Stacked structure*

A series of identical unit cells with polydomain beams were assembled into an array (Figure S1c) to form a stacked structure. The stiff support of the stacked structure was printed using a stereolithography (SLA) printer with standard white resin (FormLab3, Formlabs Inc.,

Somerville, MA, USA). Four different stacked structures (2×1, 2×2, 2×3 and 2×4) were prepared to investigate the stacking effect on architected LCEs.

**3. Force-displacement measurements**

The mechanical behavior of the unit cell was characterized by uniaxial compression for a wide range of effective strain rates from $6\times10^{-4}$ $s^{-1}$ to $6\times10^{2}$ /$s^{-1}$. This large range of effective strain rates could not be produced by a single testing system. Compression tests at low rates were performed using an Insight 5 equipped with a 100 N load cell (MTS Systems, Eden Prairie, MN, USA). The samples were secured to the top and bottom compression platens using tape (3M™ 4016 Double Coated Urethane Foam Tape 4016, 3M Company, St. Paul, MN, USA). The unit cells were compressed by 12 mm at prescribed displacement rates of 0.01, 0.1, 1 and 10 mm/s to achieve effective strain rates of $8.33\times10^{-4}$, $8.33\times10^{-3}$, $8.33\times10^{-2}$ and $8.33\times10^{-1}$ $s^{-1}$. The stacked structures with *n* layers (2×*n*, *n*=1,2,3,4), were compressed at displacement rates of (0.01, 0.1, 1 and 10)×*n*/4 mm/s, to 11×*n* mm to apply the same effective strain rates for the different stacking number..

The compression tests at intermediate strain rates were performed using the ElectroForce 3200 with a 450 N load cell (TA Instruments, New Castle, DE, USA). A 0.5 N of preload was used to secure the sample between the compression platens and the unit cells were compressed at prescribed displacement rates of 1, 10 and 100 mm/s to 12 mm, which resulted in effective strain rates of $8.33\times10^{-2}$, $8.33\times10^{-1}$ and 8.33 $s^{-1}$. Compression tests at the highest strain rates were performed using CEAST 9350 drop tower (Instron, Norwood, MA, USA), with a 50 mm diameter flat faced impactor and a 21 kN load capacity. The unit cell specimens were fixed to the bottom platen using tape and compressed at three different impact speeds 1, 4, 10 m/s, which resulted in effective strain rates of $8.33\times10^{1}$, $3.33\times10^{2}$ and $8.33\times10^{2}$ $s^{-1}$. For the 1 m/s impact speed, the total mass of the striker was set to 7.604 kg at a striker height of 51 mm. The striker height was increased to 816 mm to increase the impact speed to 4 m/s. To further increase the impact speed to 10 m/s, the striker height was increased to 1222 mm while the total striker mass was reduced to 2.604 kg.

**4. Force-displacement data for unit cells**

The mechanical response of unit cells with polydomain LCEs were characterized by force-displacement curves during uniaxial compression test (Figure S2-S4). The total energy absorption was calculated by numerically integrating the area underneath the force-displacement curve over the full range of applied displacements using the trapezoidal rule.

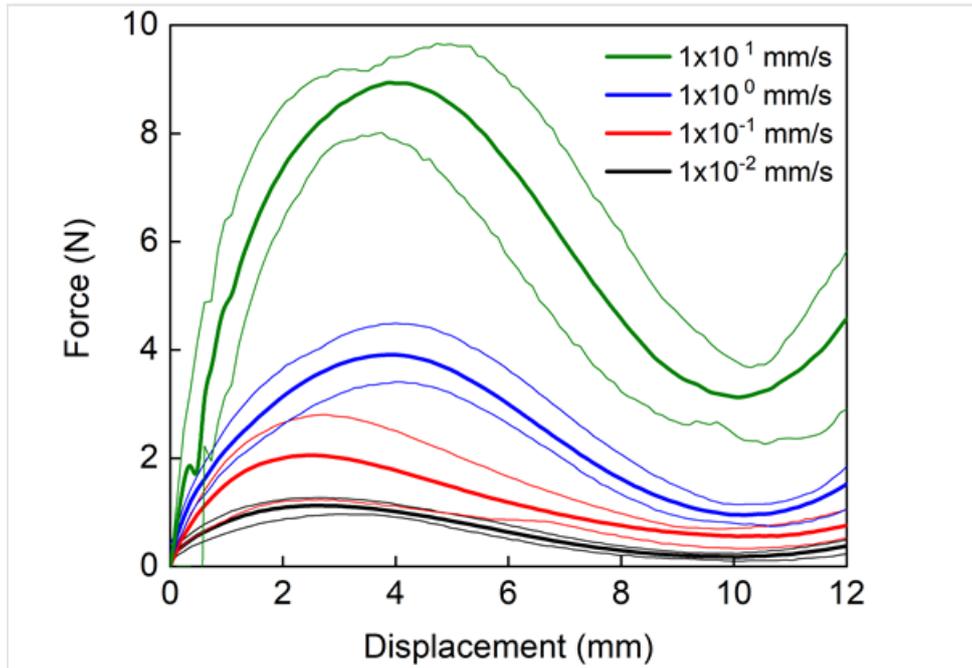

Figure S2. Experimental force-displacement curves of polydomain LCEs at different testing rates by an MTS Insight 5 (quasi-static). The thick lines indicate the average values for each testing rate and thin lines indicate the maximum and the minimum values to show the range of the data.

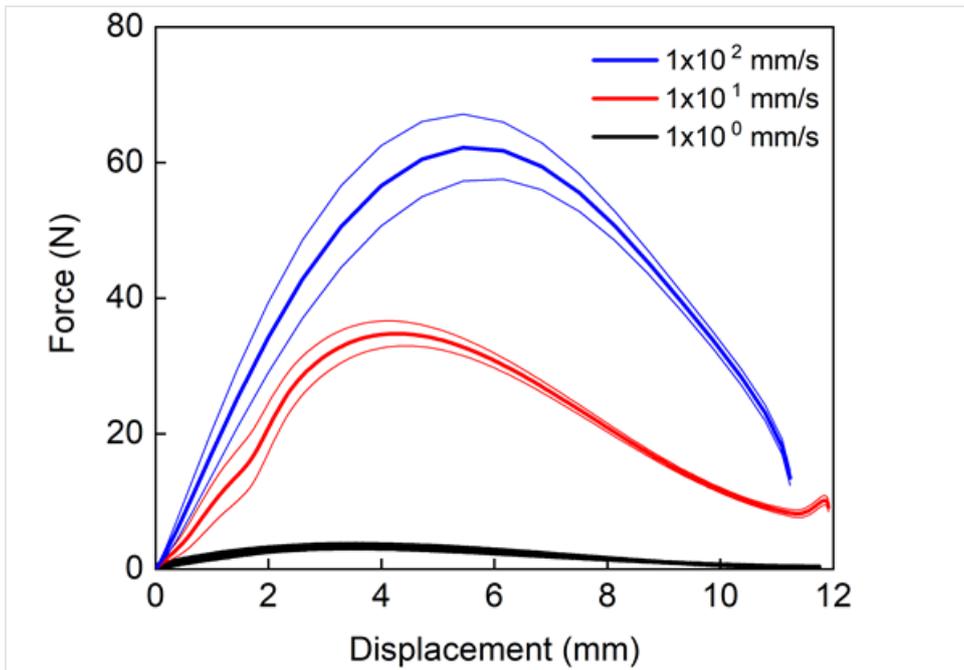

Figure S3. Experimental force-displacement curves of polydomain LCEs at different testing rates by the ElectroForce 3200 (medium rate). The thick lines indicate the average value for each testing rate while the thin lines indicate the maximum and the minimum values to show the range of data.

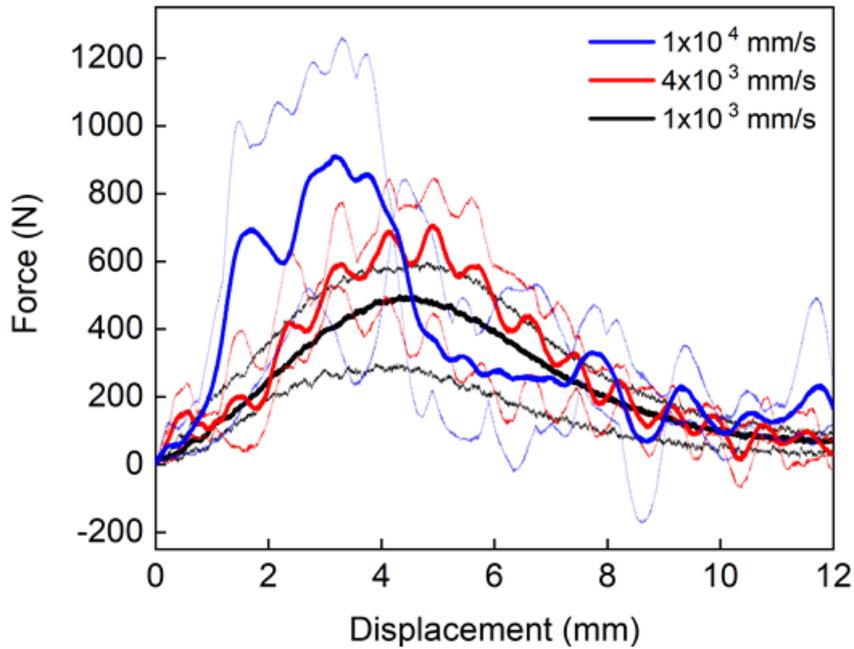

Figure S4. Experimental force-displacement curves of polydomain LCEs at different testing rates by CEAST 9350 drop tower (high rate). The thick lines indicate the average value for each testing rate while the thin lines indicate the maximum and the minimum values to show the range of data.

## 5. Force-displacement data for stacked structures

We also characterized the energy absorption capability of 2×$n$ stacked structures ($n=1, 2, 3, 4$) with polydomain beams (Figure S5-S8). The energy absorption was calculated by numerically integrating the measured force-displacement curves using the trapezoidal rule.

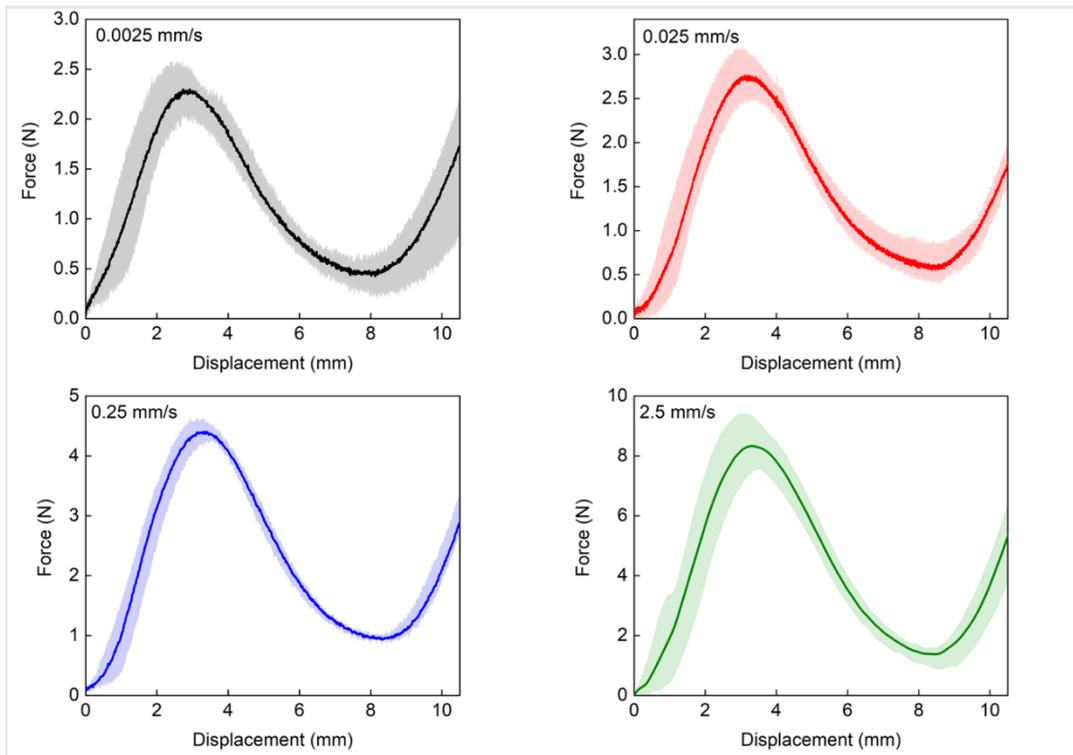

Figure S5. Experimental force-displacement curves at different testing rates for 2×1 stacked structures. Thick lines indicate the average values, and the shades indicate the data range.

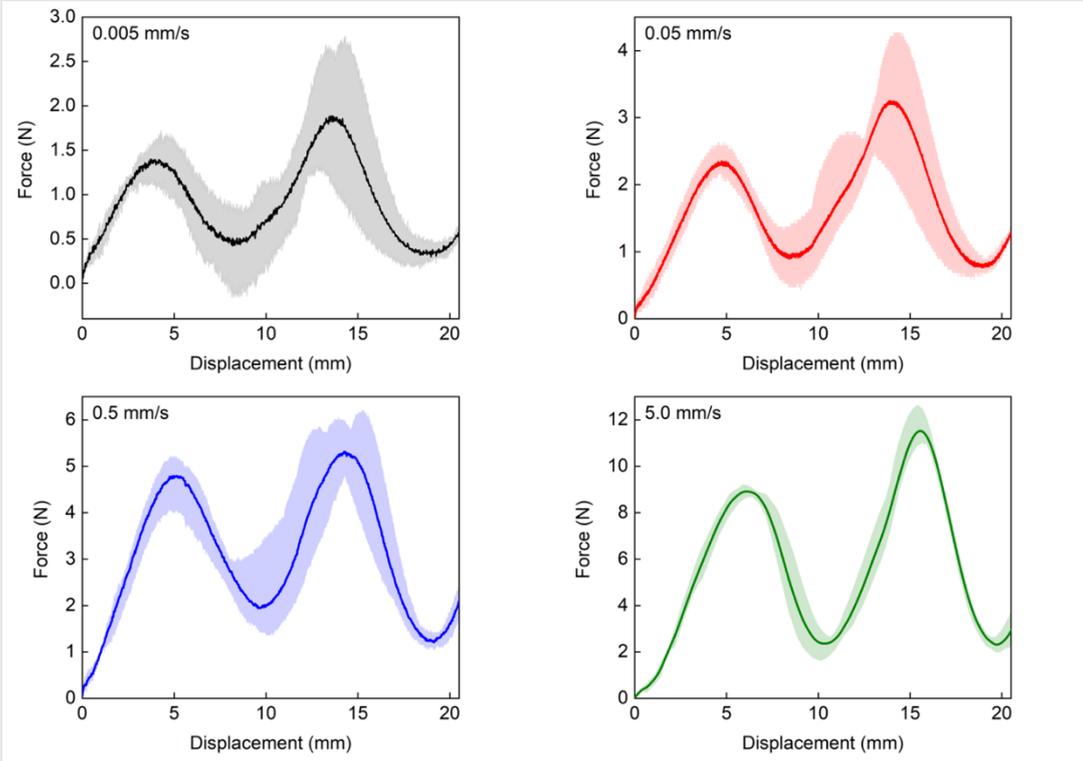

Figure S6. Experimental force-displacement curves at different testing rates for 2×2 stacked structures. Thick lines indicate the average values, and the shades indicate the data range.

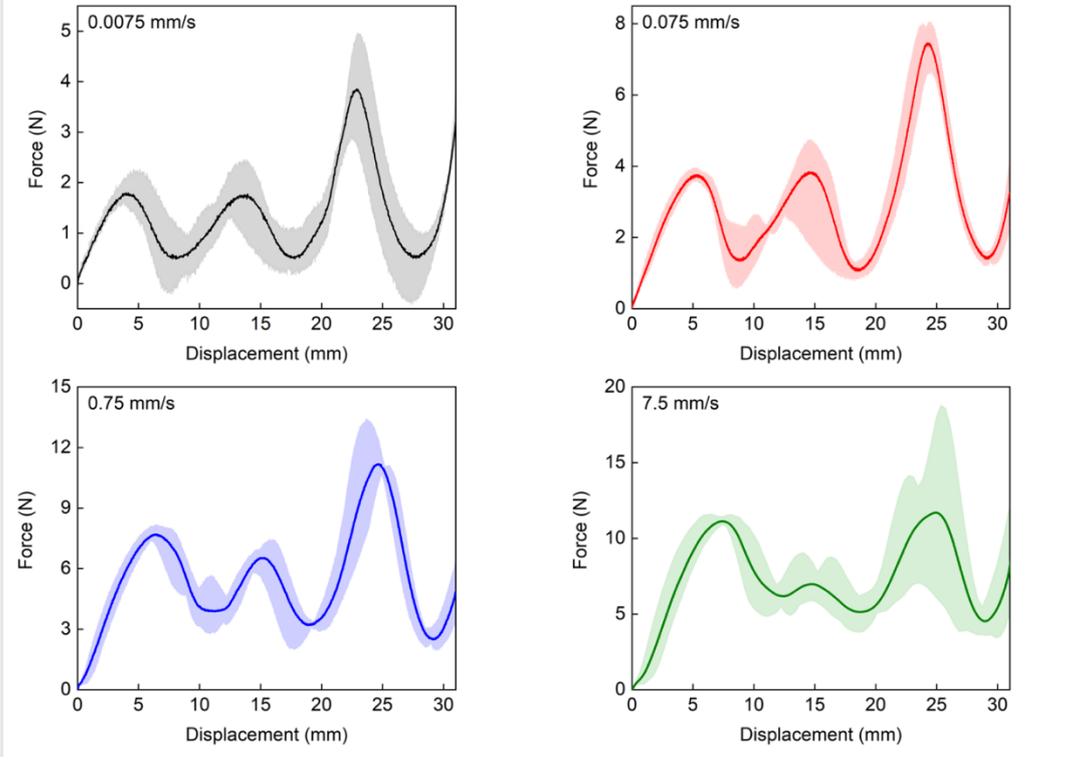

Figure S7. Experimental force-displacement curves at different testing rates for 2×3 stacked

structures. Thick lines indicate the average values, and the shades indicate the data range.

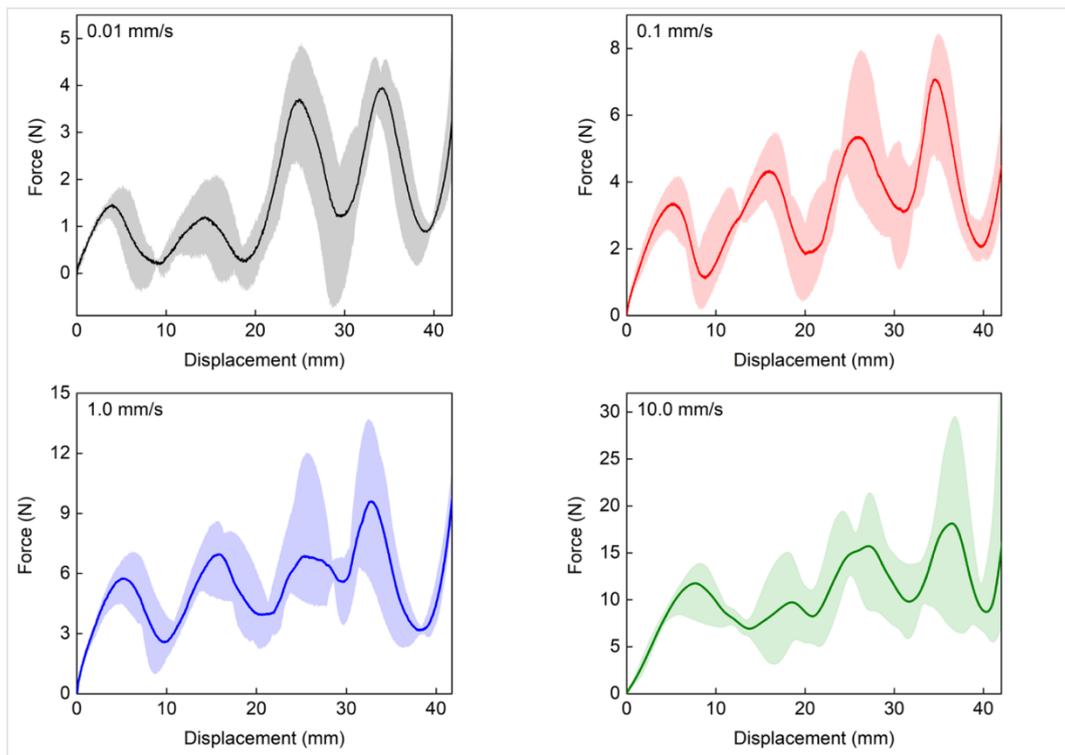

Figure S8. Experimental force-displacement curves at different testing rates for 2×4 stacked structures. Thick lines indicate the average values, and the shades indicate the data range.

## 6. Viscoelastic materials characterization

*Dynamic frequency sweep tests at different temperatures*

The viscoelastic properties of the LCE materials were characterized using a dynamic mechanical analyzer (DMA) Q800 (TA Instruments, New Castle, DE, USA) with a liquid nitrogen cooling accessory. LCE sheets were cut into multiple 25 mm by 8 mm by 1 mm strips. The specimens were mounted in tension clamps and subjected to dynamic strain of 0.1% at different temperatures and frequencies. The temperature was stepped from 0 °C to 110 °C in increments of 2.5 °C. At each temperature, the storage and loss modulus were measured for a frequency sweep from 0.02 Hz -20Hz. The master curve of storage and loss moduli for the three different LCEs (polydomain, 90% parallel (∥) and 90% perpendicular (⊥)) were constructed by horizontally shifting the storage moduli and loss moduli at different temperatures to the reference temperature of 25°C based on time-temperature superposition (TTS) principle[2] (Figure S9).

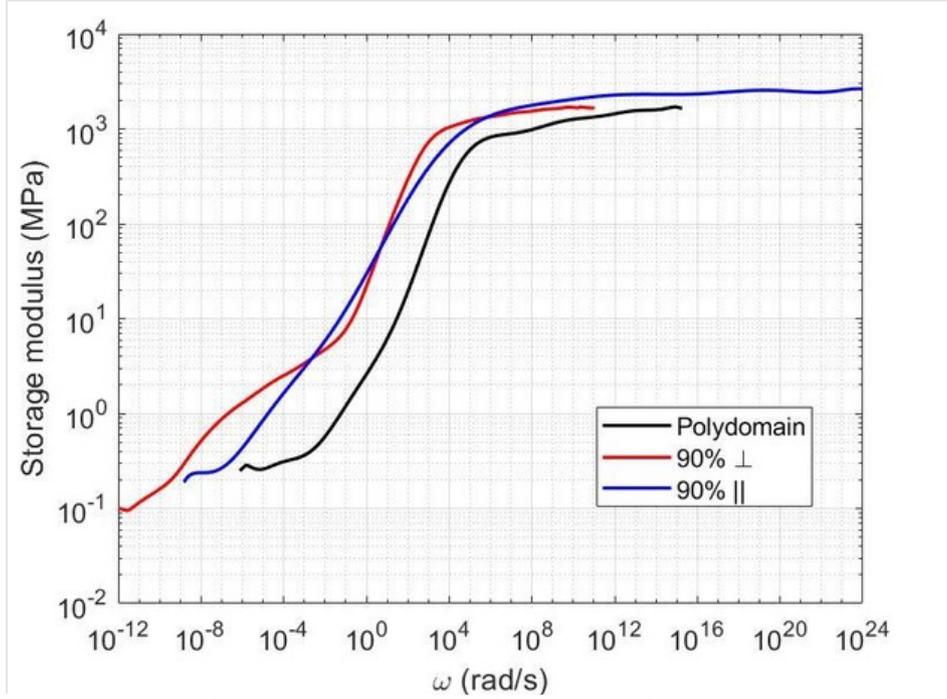

Figure S9. Master curve of the storage modulus for three different types of LCEs (polydomain, 90% parallel (∥) and 90% perpendicular (⊥)) at a reference temperature of 25°C.

*Parameter determination for viscoelastic relaxation spectrum*

The equilibrium modulus and the glassy modulus ($E^{eq}$=0.3162 MPa, $E^0$=1000 MPa) were approximated from the left and right plateau of the master curve for the truncated frequency range from $10^{-4}$ to $10^6$ (rad/s), respectively. Previous experimental studies measured a Poisson's ratio of 0.5 for the small strain behavior of LCEs[3]. Thus, the equilibrium shear modulus can be calculated as, $\mu^{eq}$= $E^{eq}$/3=0.1054 MPa. The nonequilibrium modulus is then given by $E^{neq}$= $E^0$ - $E^{eq}$=999.68 MPa. We assumed that in the glassy state, the Poisson's ratio is 0.35 such that the nonequilibrium shear modulus is $\mu^{neq}$= $E^{neq}$/2.7=615.31 MPa. A third-order approximation method developed by Schwarzl and Staverman[4] was used to determine the relaxation spectrum from the master curve of the storage modulus[5-7]. A fifteenth-order polynomial was first fit to the master curve of the polydomain LCE (Figure S9). The resulting polynomial function was expressed as

$$\log E' = f_0(\log \omega), \qquad (S1)$$

where E' is the storage modulus and ω is the frequency. The first, second, and third derivatives of the polynomial function were calculated analytically and denoted as $f_1$, $f_2$, and $f_3$, respectively. The continuous relaxation spectrum $h(\omega)$ can be approximated from polynomial function and its derivatives as follows[6]:

$$h(\omega) = 10^{[f_0(x)-x]} f_1(x) - \frac{1}{4}(f_1(x))^3 - \frac{3}{4\ln 10} f_1(x) f_2(x) - \frac{1}{4(\ln 10)^2} f_3(x)]\,|_{x=\log(\omega)}. \qquad (S2)$$

The cumulative relaxation spectrum was defined from $h(\omega)$ as,

$$H(\omega) = \int_0^\omega h(z)\, dz. \qquad (S3)$$

The storage modulus of the discrete model can be written as,

$$E'_{disc}(\omega) = 3\mu^{eq} + \sum_k^N 2.7 \mu_k^{neq} \frac{\omega^2 \tau_k^2}{1+\omega^2 \tau_k^2}, \qquad (S4)$$

where N is the number of relaxation processes, $\mu^{eq}$ is the equilibrium shear modulus, $\mu_k^{neq}$ is the non-equilibrium shear modulus of the $k^{th}$ relaxation processes at the characteristic relaxation time $\tau_k$. The cumulative spectrum $H(\omega)$ can be approximated by using a series of discrete Heaviside functions as below,

$$H_{disc}(\omega) = \sum_{k}^{N} 2.7\mu_k^{neq} <\omega - \omega_k>, \qquad (S5)$$

where $\omega_k$ is the $k^{th}$ frequency and $<\omega - \omega_k> = 0$ for $\omega < \omega_k$ ($k$=1, 2, …, N). We assumed a power law distribution for $\omega_k$:

$$\omega_k = \omega_{min}\left(\frac{\omega_{max}}{\omega_{min}}\right)^{\frac{k-1}{N-1}}, \qquad (S6)$$

where the upper and lower bounds of the frequency distribution were $\omega_{min} = 10^{-4}$ rad/s, and $\omega_{max} = 10^6$ rad/s. We defined $H_{disc}(\omega) = \frac{1}{2}[H(\omega_k) + [H(\omega_{k+1})]$, $\omega_k \leq \omega < \omega_{k+1}$, so that the discrete cumulative spectrum forms a stepwise approximation of the continuous cumulative spectrum. Then, the nonequilibrium moduli $\mu_k^{neq}$ can be determined from the continuous cumulative distribution as below:

$$\begin{aligned}\mu_1^{neq} &= \frac{1}{2.7}\left(\frac{1}{2}[H(\omega_1) + H(\omega_2)]\right), \\ \mu_k^{neq} &= \frac{1}{2.7}\frac{1}{2}[H(\omega_{k+1}) - H(\omega_{k-1})], \quad 2 \leq k \leq N-1, \\ \mu_N^{neq} &= \mu^{neq} - \sum_{k}^{N-1}\mu_k^{neq}.\end{aligned} \qquad (S7)$$

To reduce the computational time while maintaining sufficiently accurate representation of the master curve, we chose *N=16* for the discrete approximation of the relaxation spectrum. The resulting relaxation spectrum $(\tau_k, \mu_k^{neq})$ (N=1, 2, …, 16) for the different LCEs are shown in Figure S10.

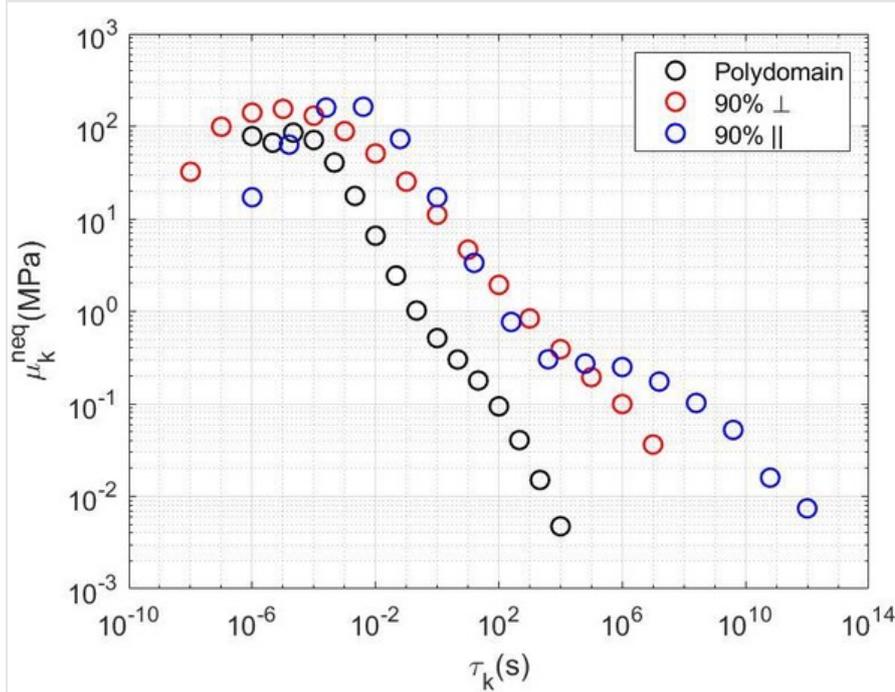

Figure S10. Stress relaxation spectrums for three different types of LCEs (polydomain, 90% parallel (∥) and 90% perpendicular (⊥)) at a reference temperature of 25°C.

### 7. Viscoelastic model

The finite deformation viscoelasticity model developed by Reese and Govindjee[9] was used to describe the constitutive behavior of the LCE materials in the finite element simulations. The model formulation is an extension of the rheological model shown in Figure S11 to finite deformation. The rheological model consists of a spring describing the equilibrium response, characterized by an equilibrium shear modulus of $\mu^{eq}$, in parallel with N Maxwell elements for the time-dependent nonequilibrium response. Each Maxwell element is characterized by a shear modulus $\mu_k^{neq}$ and relaxation time $\tau_k$.

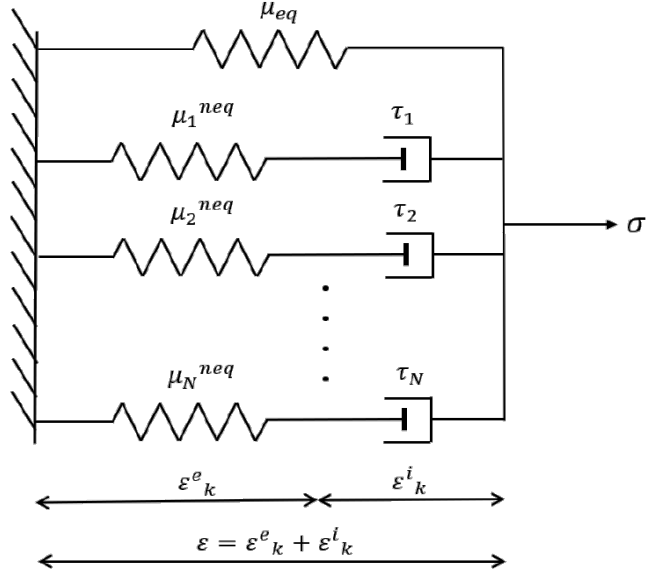

Figure S11. Rheological model for linear viscoelastic material behavior

The deformation gradient tensor is defined as $\mathbf{F} = \frac{\partial \mathbf{x}}{\partial \mathbf{X}}$, where $x(\mathbf{X})$ is a unique mapping of the reference material configuration $\mathbf{X}$ to the deformed spatial configuration. We assumed $N$ parallel multiplicative decompositions of the total deformation gradient $\mathbf{F}$ into an elastic part $\mathbf{F}_k^e$ and an inelastic part $\mathbf{F}_k^i$,

$$\mathbf{F} = \mathbf{F}_k^e \mathbf{F}_k^i, \quad k = 1\ldots N, \tag{S8}$$

where $\mathbf{F}_k^e$ and $\mathbf{F}_k^i$ are the deformation gradients associated with the $k^{th}$ spring and with the $k^{th}$ dashpot as shown in Figure S11, respectively. The right Cauchy-Green deformation tensor is defined as $\mathbf{C} = \mathbf{F}^T \mathbf{F}$, the elastic right Cauchy-Green deformation tensors are defined as $\mathbf{C}_k^e = \mathbf{F}_k^{eT} \mathbf{F}_k^e = \mathbf{F}_k^{i-T} \mathbf{C} \mathbf{F}_k^{i-1}$, and we define elastic left Cauchy-Green tensors as $\mathbf{b}_k^e = \mathbf{F}_k^e \mathbf{F}_k^{eT}$.

To model the viscoelastic behavior, it is assumed that the Helmholtz free energy density can be additively split into an equilibrium part and $N$ nonequilibrium components,

$$\Psi = \Psi^{eq}(\mathbf{C}) + \sum_{k=1}^{N} \Psi_k^{neq}(\mathbf{C}_k^e). \tag{S9}$$

For simplicity, the volumetric response is assumed to be elastic, and the free energy density is further split into equilibrium and nonequilibrium deviatoric parts and a volumetric part[11]. The Neo-Hookean model is used for both the equilibrium and nonequilibrium parts, and an Ogden model[11] is applied for the volumetric part. The resulting free energy density can be written as,

$$\Psi = \frac{\mu^{eq}}{2}\left(J^{-\frac{2}{3}}\mathbf{C}:\mathbf{I} - 3\right) + \frac{\kappa^{eq}}{4}(J^2 - 2\ln J - 1) + \sum_{k=1}^{N} \frac{\mu_k^{neq}}{2}\left(J_k^{e-\frac{2}{3}}\mathbf{C}_k^e:\mathbf{I} - 3\right), \tag{S10}$$

where $J = det(\mathbf{F})$ and $J_k^e = det(\mathbf{F}_k^e)$, and $\kappa^{eq}$ is the bulk modulus. The second Piola-Kirchhoff stress tensor $\mathbf{S}$ is defined as[9],

$$\mathbf{S} = \mathbf{S}^{eq} + \mathbf{S}^{neq} = 2\frac{\partial \Psi^{eq}}{\partial \mathbf{C}} + \sum_{k=1}^{N} 2\mathbf{F}_k^{i-1} \cdot \frac{\partial \Psi_k^{neq}}{\partial \mathbf{C}_k^e} \cdot \mathbf{F}_k^{i-T} = 2\frac{\partial \Psi}{\partial \mathbf{C}}. \tag{S11}$$

The Cauchy stress tensor can be obtained from a push forward of $\mathbf{S}$:

$$\boldsymbol{\sigma} = \boldsymbol{\sigma}^{eq} + \sum_{k=1}^{N} \boldsymbol{\sigma}_k^{neq} = \frac{2}{J}\mathbf{F}\frac{\partial \Psi^{eq}}{\partial \mathbf{C}}\mathbf{F}^T + \sum_{k=1}^{N} \frac{2}{J_k^e}\mathbf{F}_k^e \frac{\partial \Psi_k^{neq}}{\partial \mathbf{C}_k^e} \mathbf{F}_k^{eT}$$

$$= \frac{\mu^{eq}}{J} J^{-\frac{2}{3}} \left(\mathbf{b} - \frac{\text{tr}(\mathbf{b})}{3}\right) + \frac{\kappa^{eq}}{2J}(J^2 - 1) + \sum_{k=1}^{N} \frac{\mu_k^{neq}}{J_k^e} J_k^{e-\frac{2}{3}} \left(\mathbf{b}_k^e - \frac{\text{tr}(\mathbf{b}_k^e)}{3}\right), \quad (S12)$$

where $\mathbf{b}$ is the left Cauchy-Green deformation tensor, and $\mathbf{b}_k^e$ represents elastic left Cauchy-Green tensor associated with the $k^{\text{th}}$ Maxwell element.

The evolution equation for the elastic left deformation tensors is given by[9],

$$\mathcal{L}_v \mathbf{b}_k^e = -\frac{1}{\eta_k^D} J_k^e \boldsymbol{\sigma}_k^{neq} \mathbf{b}_k^e, \quad (S13)$$

where the Lie derivative of the elastic left deformation tensor $\mathcal{L}_v \mathbf{b}_k^e = \mathbf{F} \cdot \overline{\mathbf{C}_k^{i\,-1}} \cdot \mathbf{F}^T$ is an objective rate. The parameter $\eta_k^D$ is deviatoric viscosity for the $k^{\text{th}}$ dashpot in the $k^{\text{th}}$ Maxwell element and is related to the relaxation time $\tau_k$ as $\eta_k^D = \tau_k \mu_k^{neq}$.

According to the balance of mechanical energy, external work is equal to internal work. For the titled beam under compression, the external work can be written as

$$W_{\text{ext}} = \int_0^{D_y} F_y \, dD_y, \quad (S14)$$

where $F_y$ is the reaction force in the y-direction at the top of the tilted beam, and $D_y$ is the corresponding displacement in the y-direction. The internal work can be written as,

$$W_{\text{int}} = \int_0^t P_{\text{int}} \, dt = \int_0^t \int_{\Omega_0} \mathbf{S} : \frac{1}{2} \dot{\mathbf{C}} \, dV dt, \quad (S15)$$

where $P_{\text{int}}$ is the stress power. As discussed above, $\dot{\mathbf{C}} = \overline{\mathbf{F}_k^{iT} \mathbf{C}_k^e \mathbf{F}_k^i} = \dot{\mathbf{F}}_k^{iT} \mathbf{C}_k^e \mathbf{F}_k^i + \mathbf{F}_k^{iT} \dot{\mathbf{C}}_k^e \mathbf{F}_k^i + \mathbf{F}_k^{iT} \mathbf{C}_k^e \dot{\mathbf{F}}_k^i$, and

$$\left(\mathbf{S} : \frac{1}{2}\dot{\mathbf{C}}\right) = (\mathbf{S}^{eq} + \mathbf{S}^{neq}) : \frac{1}{2}\dot{\mathbf{C}}$$
$$= \left(\frac{\partial \Psi^{eq}}{\partial \mathbf{C}} + \sum_{k=1}^{N} \mathbf{F}_k^{i-1} \cdot \frac{\partial \Psi_k^{neq}}{\partial \mathbf{C}_k^e} \cdot \mathbf{F}_k^{i-T}\right) : \dot{\mathbf{C}}$$
$$= \underbrace{\left(\frac{\partial \Psi^{eq}}{\partial \mathbf{C}} : \dot{\mathbf{C}} + \sum_{k=1}^{N} \frac{\partial \Psi_k^{neq}}{\partial \mathbf{C}_k^e} : \dot{\mathbf{C}}_k^e\right)}_{\dot{\Psi}} + \left(2\mathbf{C}_k^e \frac{\partial \Psi_k^{neq}}{\partial \mathbf{C}_k^e}\right) : \left(\dot{\mathbf{F}}_k^i \cdot \mathbf{F}_k^{i-1}\right). \quad (S16)$$

Therefore, the internal work is derived as below:

$$W_{int} = \int_{\Omega_0} \Psi \, dV + \sum_{k=1}^{N} \int_0^t \int_{\Omega_0} \underbrace{\left(2\mathbf{C}_k^e \frac{\partial \Psi_k^{neq}}{\partial \mathbf{C}_k^e}\right) : \left(\dot{\mathbf{F}}_k^i \cdot \mathbf{F}_k^{i-1}\right)}_{D_{\text{visc}}} dV dt,$$

Then using $W_{\text{ext}} = W_{\text{int}}$,

$$\int_0^{D_y} F_y \, dD_y = \int_{\Omega_0} \Psi \, dV + \sum_{k=1}^{N} \int_0^t \int_{\Omega_0} \underbrace{\left[J_k^e \boldsymbol{\sigma}_k^{neq} \cdot \mathbf{b}_k^{e-1} : \left(-\frac{1}{2}\mathcal{L}_\vartheta \mathbf{b}_k^e\right)\right]}_{D_{\text{visc}}} dV dt, \quad (S17)$$

where $\Psi$ is the elastic stored energy density and $D_{\text{visc}}$ is the rate of viscous dissipated energy density. The first term on the right-hand side is the energy stored in the LCE beams while the second term is the energy dissipated by viscoelastic deformation.

## 8. Finite element models

To investigate the energy absorption mechanism of the architected LCEs beyond experiments, finite element simulations were conducted for both unit cell and stacked structures to calculate the energy dissipated by material viscoelasticity and the energy stored by snap buckling. A large deformation viscoelastic constitutive model[9] with discrete viscoelastic spectrum was applied for the LCE materials. The calculated relaxation spectrum (Figure S10) was used for the model parameters $\mu_k^{neq}$ and $\eta_k^D = \mu_k^{neq}/\tau_k$, and the material properties of the white resin were used to simulate the horizontal layers of stacked structures (bulk modulus is 4666.67 MPa, shear modulus is 1000 MPa). To avoid volumetric locking, first-order displacement and zero-order pressure scheme was used, and two-dimensional 4-node quadrilateral element was chosen throughout the simulation. For each case, the evolution of the reaction force along the vertical direction was evaluated to calculate the energy absorption, while the stored energy and dissipated energy were also calculated to compare their contribution to the energy absorption.

*Unit cell*

For the unit cell, we assumed for simplicity that the significantly stiffer support structure was rigid and the deformation was symmetric for the 2 titled beams (Figure 1b). To further simplify the model, we assumed plane strain deformation. Then, the finite element model for the unit cell was created for a single beam, tilted at an angle of 45°, and rigidly clamped at both ends. The thickness of the beam was taken from experiments as 1.8 mm, and the beam length was 13.14 mm. The beam was discretized by quadrilateral elements with element size 0.18 mm in thickness and 0.438 mm in length. To avoid volumetric locking, we used a mixed element formulation with constant pressure interpolation and bilinear displacement interpolation. Both the horizontal and vertical displacement components were fixed, $u_x=u_y=0$, at the bottom end of the tilted beam and the horizontal displacement component was constrained at the top end of the beam such that $u_x=0$ to model the rigid clamp conditions. A vertical displacement was applied to the top end of the beam, $u_y=vt$ (Figure S12). The displacement rate $v$ was set to the 10 mm/s, 1 mm/s, 0.1 mm/s and 0.01 mm/s applied in the low-rate experiments (Figure 2c).

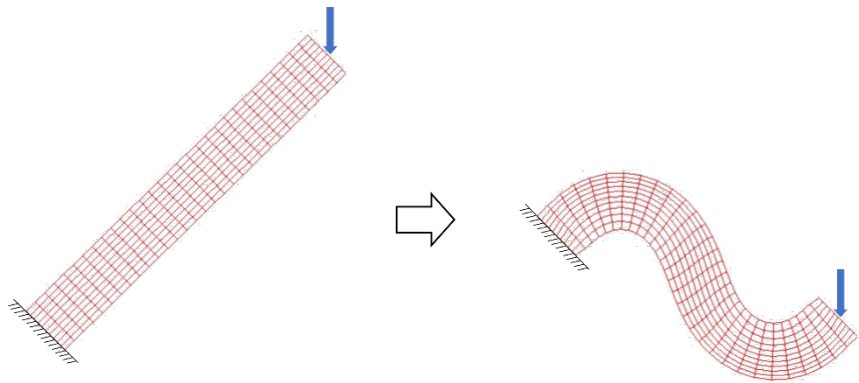

Figure S12. Finite element model for LCE beam

The reaction force in the vertical direction was calculated at the node with the applied displacement to calculate the energy absorption according to Eq. 14 (Section 7). The element size was chosen by a convergence study for the reaction force. The force was normalized by peak force for the fastest loading rate, 10 mm/s, for experiments and simulations, and plotted as a function of displacement for different loading rates (Figure S13). The shaded areas indicate the upper and lower bound of 3 specimens tested at each rate. The force in the simulations was generally larger than in experiments. However, the rate-dependence of the normalized force agreed well between displacements and simulations throughout the snap-through buckling event. The energy absorption density calculated using Eq. 14 (Section 7) for different loading rates was plotted in Figure S14 comparing experiments and simulations. Also plotted are the stored energy density and viscous dissipation energy density calculated using Eq. 17 (Section 7) for the simulations. The rate-dependence of the energy absorption density agreed well between experiments and simulations, but the simulations overestimated the energy absorption. Both the stored energy and viscous dissipation increased with rate in the range of the low-rate experiment (Figure 2c). The rate-dependence of two curves crosses over at the critical strain rate $6×10^{-3}$ s$^{-1}$, which indicates that energy storage was the dominant energy absorption mechanism below the critical strain rate while viscous dissipation was the dominant mechanism above the critical strain rate.

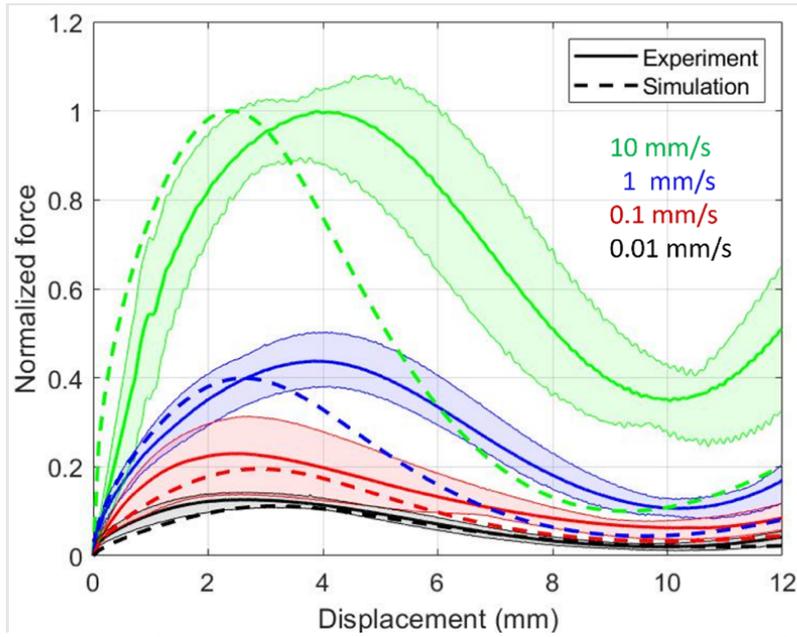

Figure S13. Normalized force-displacement curves (divided by the peak force at 10 mm/s) of polydomain LCE unit cell at different strain rates. Dash lines are simulation whereas the solid lines are experiment. The shaded areas are the upper and lower bounds of experimental data.

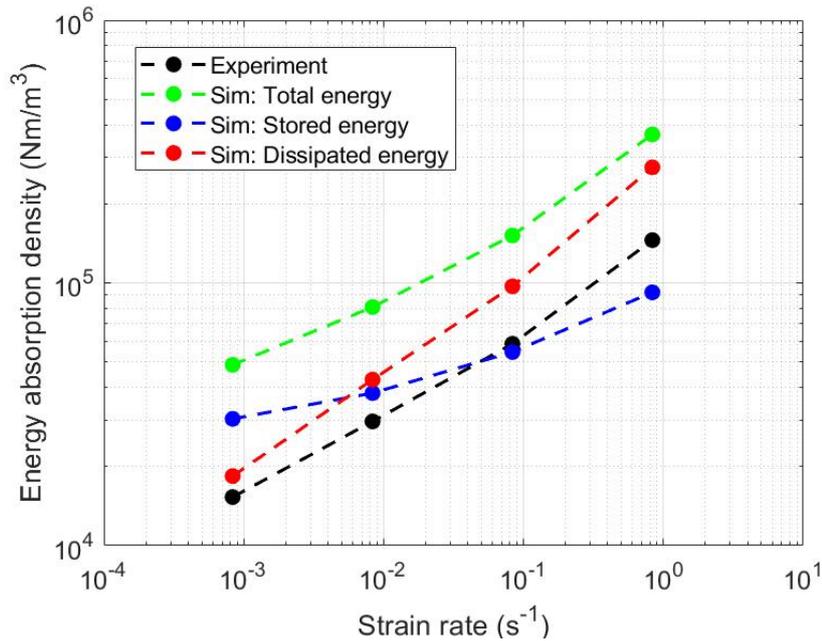

Figure S14. Energy absorption density (total energy, stored and dissipated energy) of polydomain LCE unit cell at different strain rates.

*Stacked structures*

*i) Uniform beam thickness*

Finite element simulations were performed for the 2×1, 2×2, 2×3, 2×4 and 2×8 stacked structures assuming plane strain deformation. To accurately capture the interaction of the repeating units in the structures, we modeled the entire structure, including the stiff supports (e.g., Figure S15). The same viscoelastic model and model parameters used for the polydomain LCE beam in the model of the unit cell were used for the simulations of the stacked structures. A compressible Neo-Hookean model was used to describe the behavior of the white resin material of the support structure. The bulk modulus, 4666.67 MPa, and shear modulus, 1000 MPa, were used for the resin material so that the Young's modulus of the resin material is 9000 times larger than the equilibrium Young's modulus of the polydomain. The LCE beams were titled at a 45° angle, and were 1.05 mm in thickness and 8.50 mm in the length. The displacement of the bottom boundary of the stack structure was fixed. The displacement of the top boundary was fixed in the $x$ direction, and a displacement $u_y=vt$, was applied in the direction. The displacement rate $v$ was set to be the same as in the experiments. The lateral boundaries were unconstrained on either side. Contact elements were used to prevent the beams from interpenetrating the support structure.

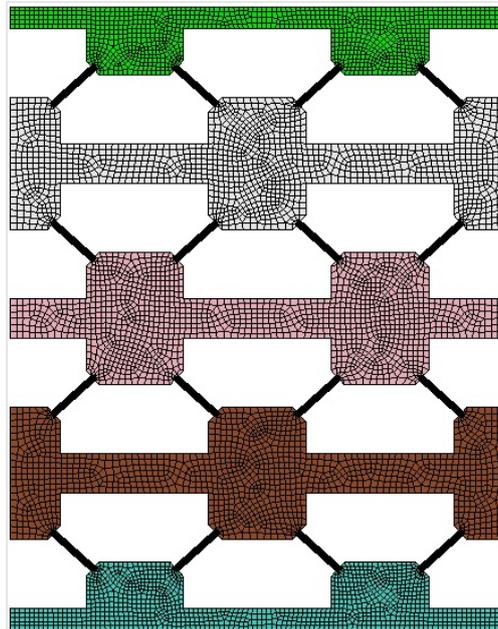

Figure S15. Geometry and mesh configuration of 2×4 stacked structure.

The reaction force was calculated for the top surface where the displacement was applied. Figure S16 plots the reaction force vs. applied displacement for the different strain rates. For structures with 3 or fewer layers, the LCE beams did not buckle uniformly (see for example. Figure 4a). The buckling of one layer causes the adjacent cells to recover (i.e., become less compressed) before compressing again once the buckled cells have fully collapsed. The sequential buckling of the layers produced multiple peaks in the force-displacement curve (Figure S16). The number of peaks corresponds with the stacking number in the stacked structure. The maximum force for each peak in magnitude because of the individual buckling of each layer. For structures with greater than 3 layers and higher displacement rates, adjacent layers began to buckle together (Figure S24) resulting in a large first peak of the force-displacement curve, and subsequent peaks became less distinct (Figure S16). For example, for the 2×3 stacked structure at a strain rate of 2.27×10$^{-1}$ s$^{-1}$, two adjacent layers buckled together resulting in a large first peak. The force displacement curve exhibited only 2 peaks instead of 4 (Figure S16). The viscoelastic behavior of the beams and the nonuniform strains produced by beam bending likely induced variations in the stress state of the beams that caused sequential buckling. It is unclear why the beams buckle together for higher stacking number $n=3$.

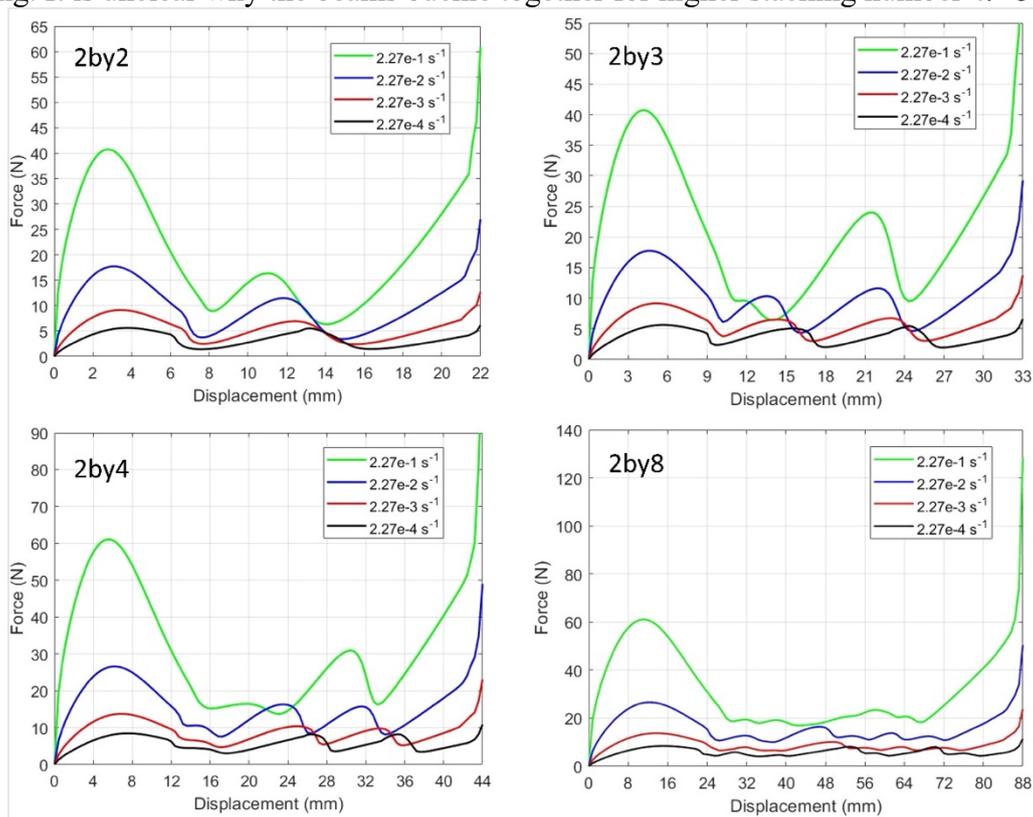

Figure S16. Force-displacement curves of 2×2, 2×3, 2×4 and 2×8 stacked structures at four different strain rates.

The energy absorption density is plotted as a function of strain rate for the different (*2×n*) stack structures in Figure S17 and as a function of stacking numbers *n* for different strain rates in Figure S18. For all strain rates, the energy absorption density increases with stack number *n* up to 3 stack layers, which shows that sequential buckling of the layers produced more energy absorption. When a layer buckles, the adjacent layer recovers then compresses again. This cyclic behavior produces additional viscous dissipation. Figure 19 plots the stored energy density and the viscous dissipation density as a function of strain rate for the different stacked structure. The stored energy density was identical for the different stacked structures. However, the viscous energy density increased with the stacking number up to *n=3*.

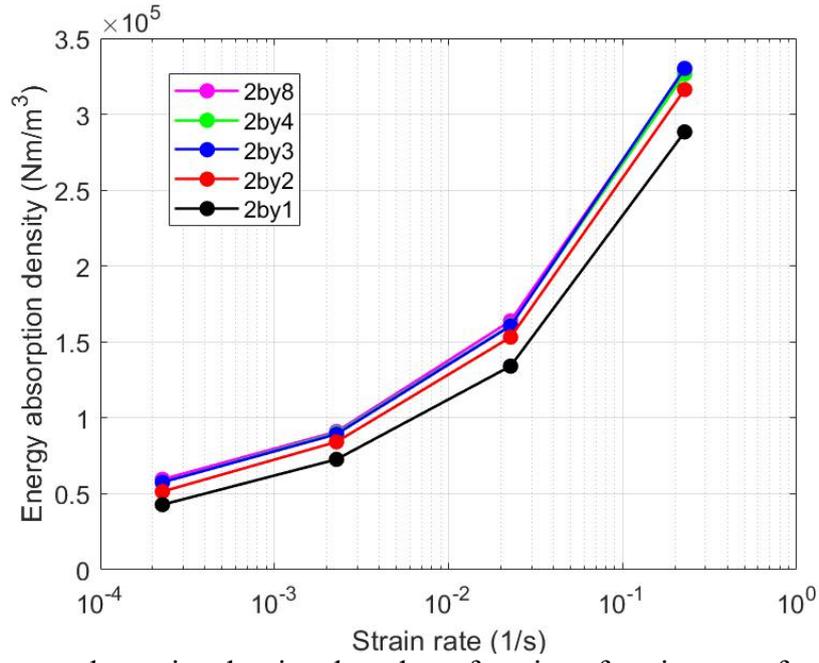
Figure S17. Energy absorption density plotted as a function of strain rate at for different stacked structure.

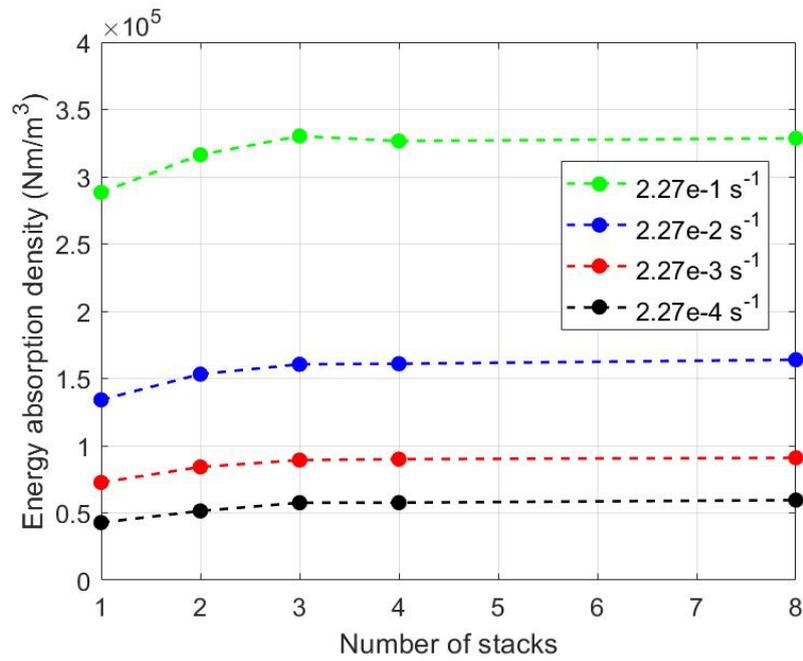
Figure S18. Energy absorption density plotted as a function of stacking number at four different strain rates.

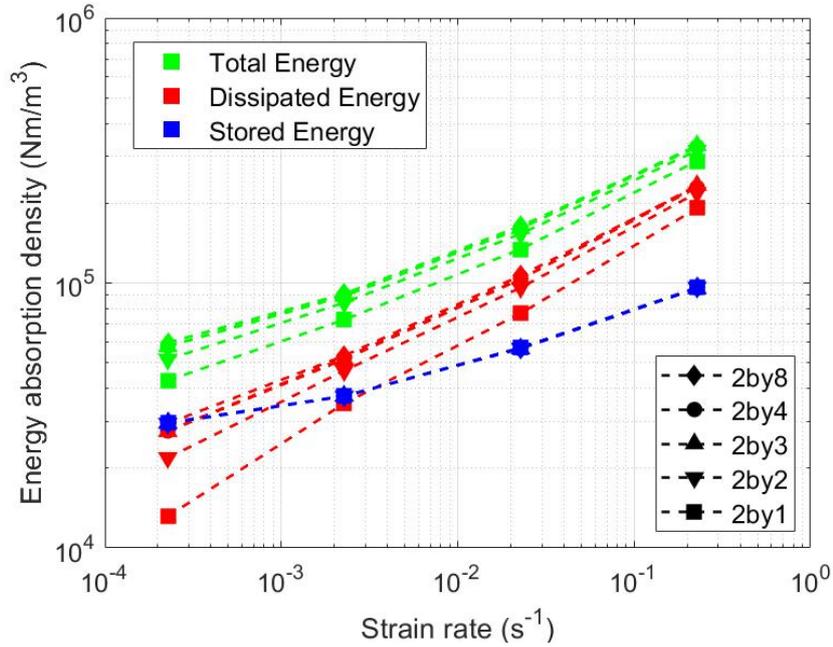

Figure S19. Energy absorption density (total energy, stored and dissipated energy) plotted as a function of strain rates for different stacking number.

Figure S20 plots total energy absorption for the different stacked structures to further illustrate the synergy between snap buckling and viscoelastic dissipation. If the stacked structures were elastic, the total energy absorption would increase linearly with n because the energy absorption density from snap buckling is independent of stacking number. In contrast, the energy absorption for the stacked structures with the LCE beams increased nonlinearly because of the increased viscous dissipation caused by sequential snap buckling. The shaded areas indicate the additional total energy absorption above the linear increase.

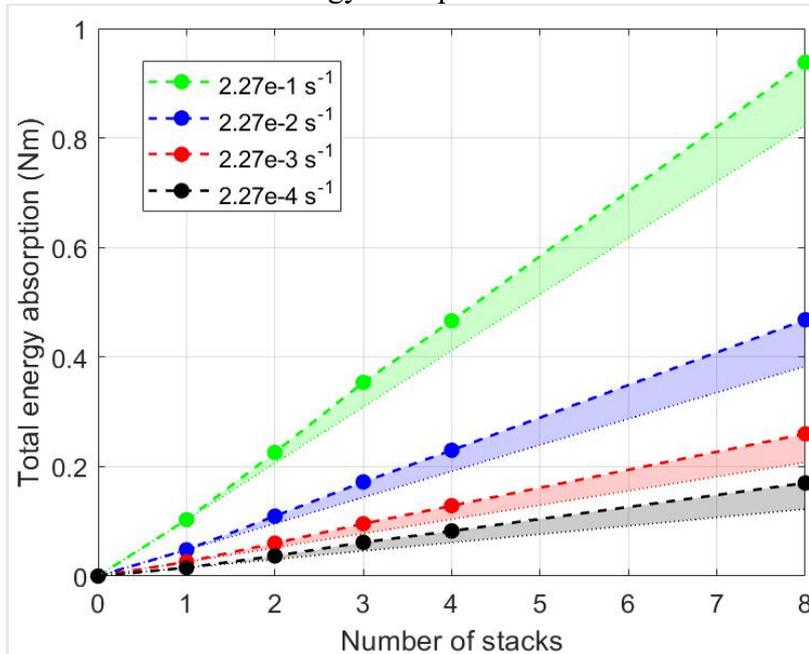

Figure S20. Total energy absorption plotted as a function of stacking number at four different strain rates. The dash lines are the fit to the data, and the thinner dotted lines are the reference assuming that the total energy absorption is proportional to stacking number if it is elastic. The shaded areas are the additionally non-linear incremental total energy absorption of stacked structures with polydomain LCE.

*ii) Graded stacked structures*

Finite element simulation of the stacked structures showed that non-uniform buckling of the different layers produced increased viscous dissipation and increased energy absorption. However, the phenomenon was limited to structures with up to 3 layers. As the stacking number increased, the adjacent layers would increasingly buckle together until the stacked structure buckled uniformly. To promote non-uniform buckling in the higher number of stacked structures, we increased the thickness of the beams by 0.05 mm from the preceding layer while keeping the total volume of all the beams the same as for the uniform structure. Thus, for a 2×8 stacked structure, the thickness of the beam at the top layer and bottom layer were 0.875 mm and 1.225 mm, respectively. The force-displacement curves for the different stack structures are plotted in Figures S21-S23 comparing the structures with a uniform beam thickness and structures with a linear variation in beam thickness. Varying the beam thickness produced sequential buckling of the layers for all stack structures and all rates (Figure S24). The force-displacement curve exhibited the same number of peaks as the stacking number and each peak exhibited the similar force. The stored and dissipated energy density was plotted in Figure S25. The structures with the varying beam thickness dissipated more energy than structures with uniform beam thicknesses, and the difference was magnified at higher strain rates. As expected, the viscous dissipation energy density difference between non-uniform and uniform cases increased with the stacking number, while the stored energy density difference did not vary significantly. Moreover, unlike the uniform cases, the energy density of the graded structures continued to increase with stacking number without saturation (Figure S26) because the mechanism of sequential buckling of the layers occurred regardless of the number of layers or applied displacement rate.

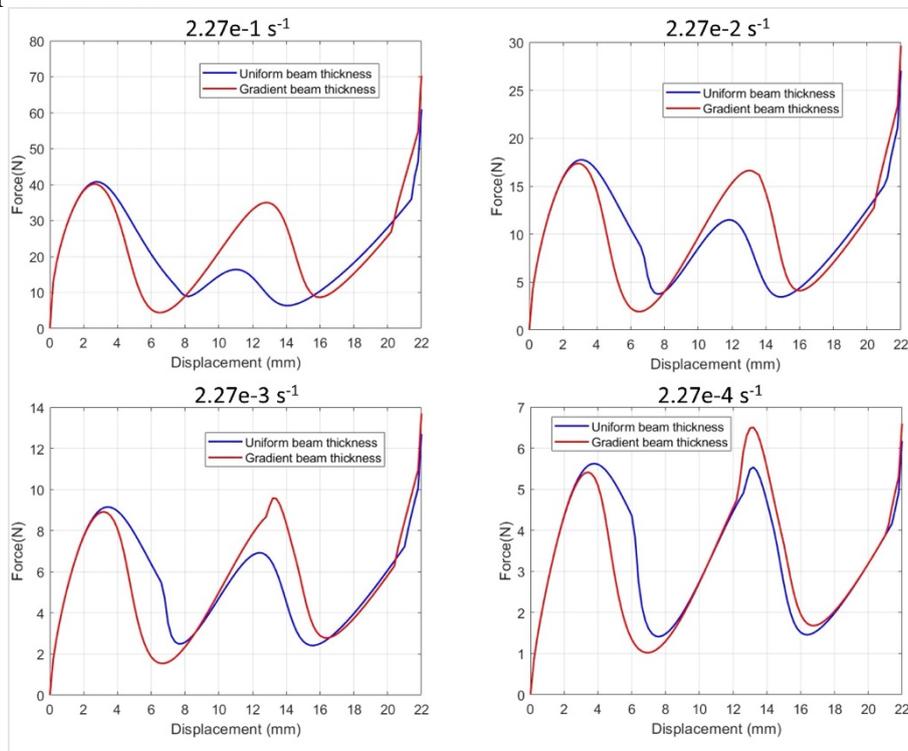

Figure S21. Force-displacement curves of both uniform and graded beam thickness cases for 2×2 structures at different strain rates.

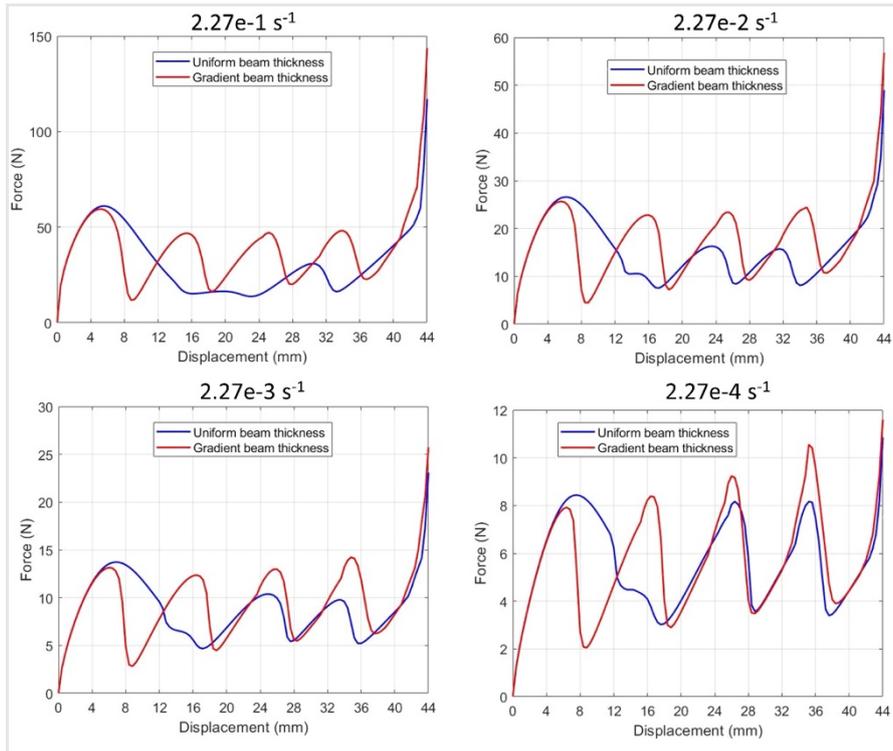

Figure S22. Force-displacement curves of both uniform and graded beam thickness cases for 2×4 structures at different strain rates.

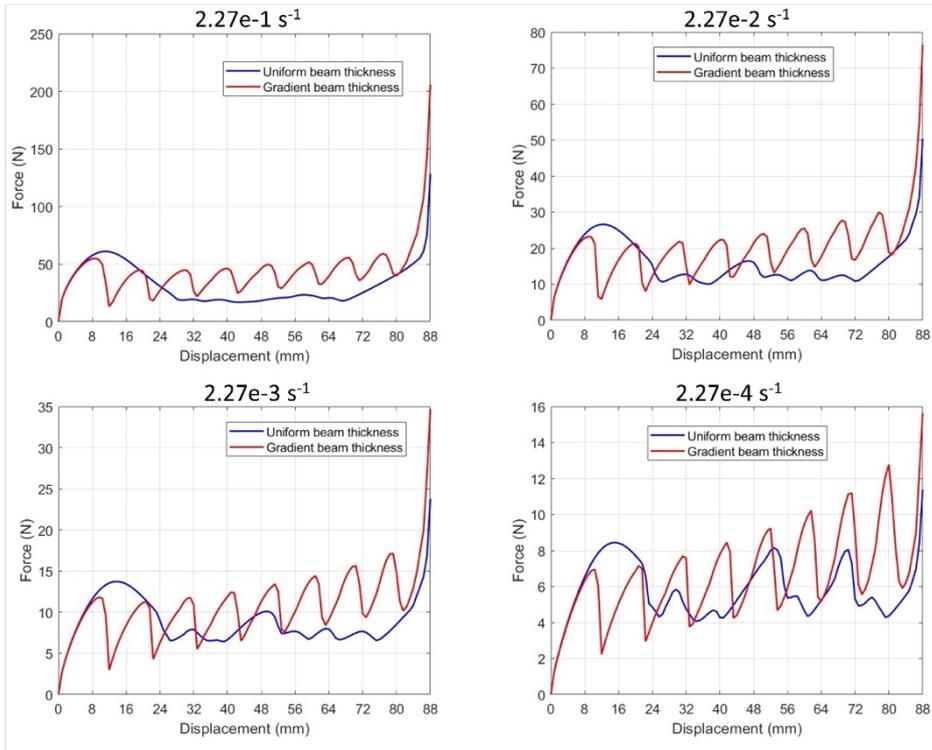

Figure S23. Force-displacement curves of both uniform and graded beam thickness cases for 2×8 structures at different strain rates.

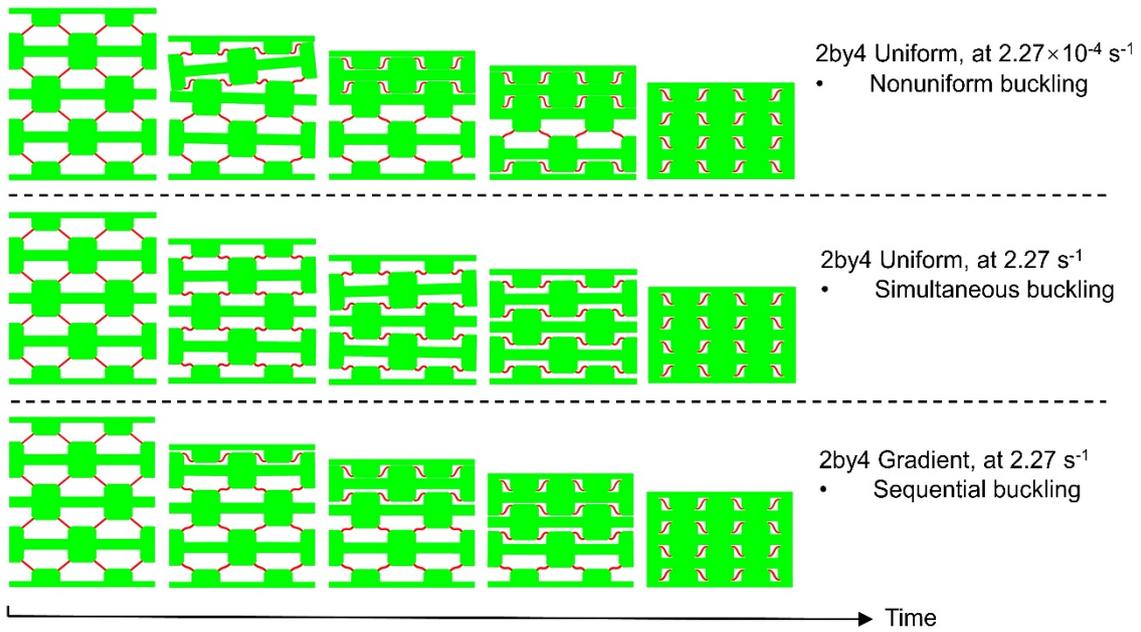

Figure S24. Deformation snapshots of 2 by 4 stacked structures for both uniform and graded cases at strain rates of 2.27 ×$10^{-4}$ $s^{-1}$ and 2.27 $s^{-1}$.

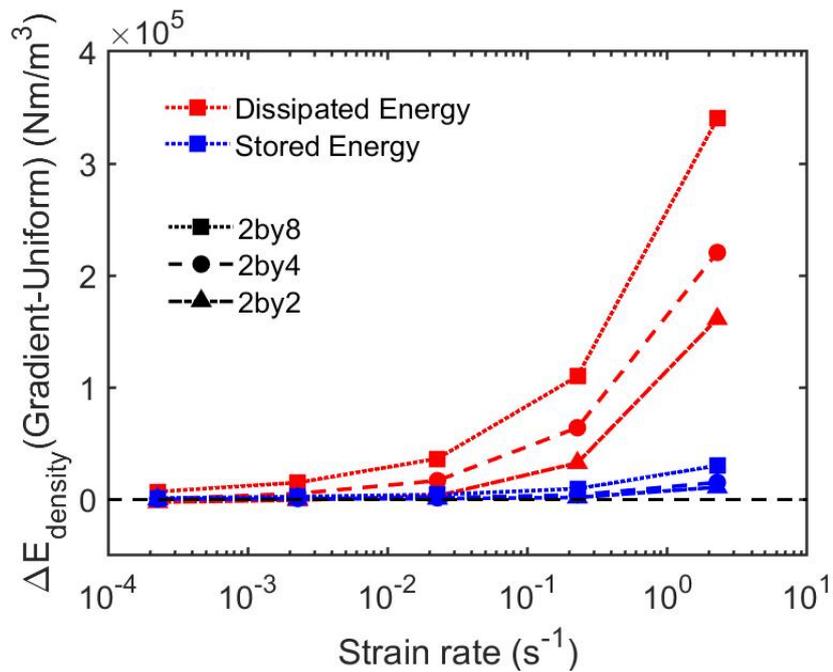

Figure S25. Stored and dissipated energy density difference between graded and uniform cases for 2×2, 2×4 and 2×8 stacked LCE structures plotted as a function of strain rates. The horizontal dash line passing through zero indicates the case where energy density of graded case is equal to that of uniform case.

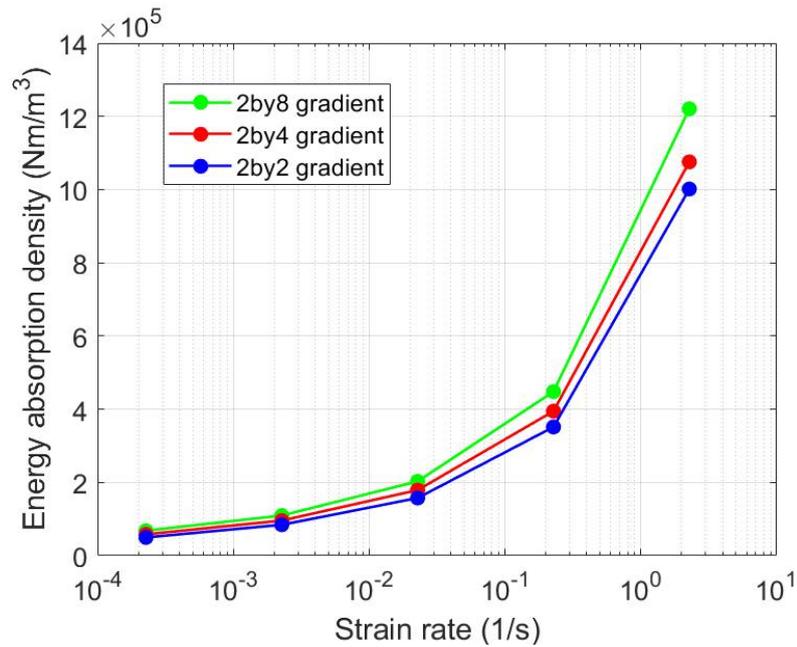

Figure S26. Energy absorption density of 2×2, 2×4 and 2×8 stacked LCE structures with graded beam thickness plotted as a function of strain rates.

## 9. Movies
Movie S1: A movie of a high strain rate impact test (impact speed = 10 m/s).
Movie S2: An experimental movie showing sequential deformation and partial recovery.
Movie S3: Simulation movies comparing uniform vs. graded cases of stacked structures.

**Supplemental References**